\pdfoutput=1
\documentclass[prl,aps,twocolumn,groupedaddress,super]{revtex4}
\usepackage{amsmath}
\usepackage{amsbsy}
\usepackage{amssymb}
\usepackage[dvips]{graphicx}
\usepackage{epsfig}
\usepackage{color}

\begin{document}
\title{
Edge states mechanism for the anomalous quantum Hall effect in 
diatomic square lattice }
\author{B. Ostahie, M. Ni\c t\u a, and A. Aldea}
\affiliation{ National Institute of Materials Physics, POB MG-7,
77125 Bucharest-Magurele, Romania} 
\date{\today}
\begin{abstract}

The understanding of the Chern insulator and   anomalous quantum
Hall effect (AQHE) in terms of chiral edge states in confined systems 
is the first aim of the paper. The model we use consists in a diatomic 
square lattice with hopping to the next-nearest-neighbors and broken 
time-reversal symmetry, which  exhibits  edge states  in the absence 
of an  external magnetic field.
The question of chiral edge states is approached in the  ribbon and plaquette 
geometries  with different atomic connectivities at the boundaries. 
Insulating and semimetallic phases are revealed, the resulting phase diagram
being richer than  in  Haldane's model. 
The transmission coefficients and  the Hall resistance $R_H$ are calculated 
for the finite size system in the Landauer-B\"{u}ttiker formalism. 
The quantized values $R_H=\pm h/e^2$, specific to the Chern insulator, 
are manifest in the energy range  occupied by the chiral edge states, 
corresponding to the unique gap existing in the energy spectrum. 
Our second aim is to examine the disorder-induced properties of the model, and, 
as a novelty, we prove the {\it disorder-driven} AQHE in the semimetallic phase.  

\end{abstract}
\maketitle
\section{Introduction }
The paradigm of the anomalous quantum Hall effect  was advanced by 
Haldane  for a hexagonal lattice model by imposing two 
{\it sine qua non} conditions: i) the hopping process to the 
next-nearest-neighbors (n.n.n), and ii) the phase attached to this process. 
The phase is justified in terms of a periodic internal magnetic field, 
chosen such that the flux through the  unit cell vanishes.
The phase diagram of  the model contains domains with  topological properties, 
where the  Chern number is $\nu=\pm1$.  Such systems, which may 
support quantum Hall effect in the absence of an external magnetic 
field, have been called Chern insulators. 
The topic  has been developed next to account for the robustness  of the effect
in the presence of disorder \cite{EVCastro,Prodan}, and for the 
interactions in the fractional anomalous quantum Hall effect \cite{Neupert,Sheng}. 
The same model has been  extended by Kane and Mele by adding the spin-orbit 
coupling and proving theoretically  the quantum spin Hall effect \cite{KM}.
The case of the edge states in the Haldane's  honeycomb model has been invoked in 
\cite{Hao,Fradkin}.

In principle, by virtue of the  bulk-edge correspondence \cite{Hatsugai},
alternatively to the approach based on topological invariants 
(which assumes Bloch functions and works for {\it infinite} systems), 
the topological properties can  be  addressed   in terms of 
chiral edge states existing in {\it finite} systems. 

This paper aims to detect chiral edge states, assumed to exist 
at vanishing magnetic flux under specific conditions and to support 
anomalous  quantum Hall effect in  Chern  insulators. 
The confirmation of such  states provides  more physical insight into 
this effect, and it is  by itself an interesting issue that certifies 
the bulk-edge correspondence also in this case.

Next, the paper observes the presence of a semimetallic phase 
where edge states  are intercalated with bulk states.
By the use of disorder, we localize the bulk and activate
the edge states, which prove to be robust and to support quantized
transport. This result advances the new  conceptual issue  of the 
{\it disorder-driven} anomalous quantum Hall effect. 

The diatomic square lattice with nearest-, and next-nearest-neighbor 
hopping \cite{Hou} is a  rich model exhibiting different phases:
semimetallic (SM), band insulating (BI), and Chern insulating (CI). 
Recently, this lattice was also considered to advocate  the presence of
spin-polarized flat bands in topological crystalline insulators \cite{Sessi}.
The model consists of two square lattices built of atoms $A$ or $B$, 
respectively, interconnected by a complex  nearest-neighbor hopping term 
$t_{AB}=t_1e^{i\gamma}$. 
So far the lattice is bipartite, however the model is completed 
with the  next-nearest-hopping $t_{AA}=t_{BB}=t_2$ that 
breaks this property by connecting atoms of the same type.

Being known that the bipartite symmetry of a lattice ensures the 
electron-hole symmetry of the energy spectrum \cite{Mielke}, the present model 
is an  unexpected  example that exhibits the spectral symmetry
in the absence of the lattice bipartitism (see Eq.(8)).

The breaking of the time-reversal (TR) symmetry, necessary for the 
appearance of the Hall effect, comes from  the phase $\gamma$ attached
to the n.n. hopping parameter $t_1$. 
It turns out, however, that the condition  $\gamma \ne 0$ by itself 
is not sufficient to get topological states, and the presence of 
the n.n.n. hopping parameter $t_2\ne 0$ is also necessary.

Different  phases  (SM,BI,CI) can be identified at once from the expression 
of the energy spectrum Eq.(8), which is  obtained easily using the 
Fourier transform of the Hamiltonian describing the infinite lattice.
Beside $t_1,t_2$ and $\gamma$, a fourth parameter  proves to be important, 
namely the energy spacing  $\Delta=(E_a-E_b)/2$ (where $E_a,E_b$ are 
the site energies in the Hamiltonian (1)). 

The evidence of the edge states is more conveniently studied  in the 
ribbon (strip) geometry, in which case  analytic calculations are
also accessible. On the other hand, the study of the transport properties, 
and, in particular, the calculation of the Hall conductance 
in the topological phase, requires a plaquette geometry.

The ribbons are tailored by cutting  the edges along different  directions 
in the lattice. Since the  edges may show different atomic connectivities, 
the ribbons may also differ in their spectral properties.
Edges parallel to A-A and A-B directions in the diatomic lattice (see Fig.1) 
are considered in Sec.III. The location of the edge  states along the 
ribbon boundaries is attested analytically and numerically. 

The ribbon energy spectrum may be gapped, corresponding  to  conventional 
(bulk) insulators, if the gap is empty, or to topological insulators, 
if the gap is filled with edge states. 
A semimetallic phase containing edge states is also found.
In this case, depending on parameters, the edge states  are
organized  as a flat band located in the middle of the spectrum 
(similar to the zig-zag graphene ribbon), or may be dispersive.
The second case is more exotic. The  point  is that the dispersive edge states 
are embedded in the  continuum of the bulk states, as  evident in 
Fig.5a and Fig.5b. The issue of the  disorder-driven topological phase, 
discussed in Sec.IV, is based on this observation.

The finite plaquette is obtained by imposing vanishing boundary 
conditions all around the perimeter.
The description of the energy spectrum as function of the  phase 
$\gamma$ is the analogue of the Hofstadter-type  spectrum  for the 
confined 2D electron gas, expressed in terms of the external magnetic flux. 
Although both spectra contains edge states, there are striking 
differences. The Hofstadter butterfly contains a sequence 
of bands and gaps originating in the Landau levels. By imposing  
boundaries, the gaps get filled with  edge states, whose chirality 
is given by the derivative of the eigenenergies with respect 
to  the magnetic flux.
In contrast, the energy spectrum  of the finite square diatomic 
lattice contains a single gap, which accommodates edge states and exhibits 
a 'binocular' aspect (see Fig.9b). The proof of chirality should be  
addressed however differently \cite{Note}, and it can be  done either by 
calculating the diamagnetic moments associated to the edge states or 
showing the asymmetric behavior of the transmission coefficients in 
the Landauer-B\"{u}ttiker approach \cite{LB}.
The second route will be followed in Sec.IV of the paper.

The edge states embedded in the multitude of bulk states, discovered first 
for the ribbon geometry, deserves special  attention 
in the case of plaquettes. Being masked by the bulk states, 
they cannot be observed in the transport, and, in order to 
evidentiate their contribution, one has to find the proper tool 
to obstruct the states in the bulk. The question is  addressed in  Sec.IV. 
in the frame of disorder induced localization.

The surprising result is that, by increasing gradually the disorder, 
the quantized Hall resistance $R_H=\pm h/e^2$ becomes manifest, 
as shown in Fig.12.
This outcome demonstrates that the edge states survive to disorder, 
while the states in the bulk get localized, being unable to participate 
in the transport.
This proves that the  novel edge states, identified in the 
semimetallic phase, are chiral  topological states, 
%which support the anomalous quantum Hall effect.
able to  support the anomalous quantum Hall effect.

The transport calculations are performed numerically in the 
Landauer-B\"{u}ttiker approach \cite{LB}, which pretends the knowledge of 
the Green function corresponding to the full Hamiltonian describing the 
sample, the leads and the coupling between them.

The model Hamiltonian, its symmetries, and possible phases 
are detailed in Sec.II. Section III is devoted to spectral properties  
in the ribbon geometry, with focus on the edge states. 
The spectral and transport properties of the clean
and disordered plaquettes, with focus on the Hall resistance quantization  
in the Chern insulating phase are discussed in Sec.IV. The conclusions  are
summarized in the last section.

%Fig1
\begin{figure}
\includegraphics[angle=-00,width=0.30\textwidth]{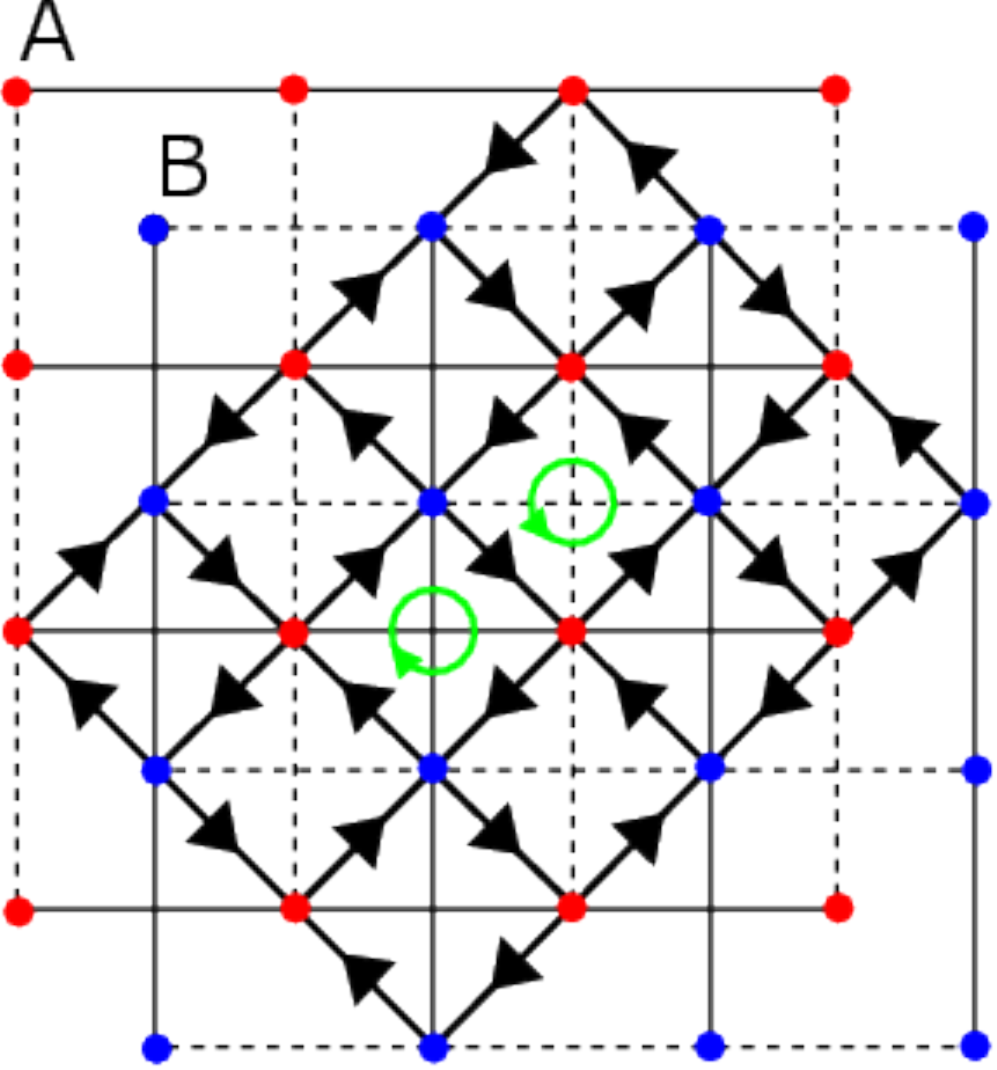}
\caption{(Color online) Diatomic square lattice structure: 
A and B atoms are interconnected by the hopping $t_1$ carrying 
the phase $\gamma$ (represented by black arrows). 
The green arrows indicate the phase loops in the lattice.  
The A-A and B-B hopping parameter $t_2$ is real, but taken with 
positive/negative sign along the solid/dashed lines.} 
\end{figure}

\section{The Hamiltonian, symmetries and spectral properties }
We consider the square lattice with two atoms A and B per the
unit cell defined by the primitive vectors  $\vec{a}_1,\vec{a}_2$.
Let us introduce the creation (annihilation) operators 
$a_{nm}^{\dag},b_{nm}^{\dag}
(a_{nm},b_{nm})$, corresponding to the atoms in the unit cell
specified by the lattice vector $\vec{R}_{nm}=n\vec{a}_1+m\vec{a}_2$.

We shall analyze the spectral and topological transport properties  
of systems described by the following tight-binding model Hamiltonian \cite{Hou}:
\begin{eqnarray}
H=\sum_{n,m}[E_a a_{nm}^{\dag} a_{nm}+E_b b_{nm}^{\dag} b_{nm}] 
~~~~\nonumber\\
-t_1\sum_{n,m}[e^{-i\gamma} a_{nm}^{\dag}(b_{nm}+b_{n-1,m-1})~~\nonumber\\
+e^{i\gamma}a_{nm}^{\dag}(b_{n,m-1}+b_{n-1,m}) + H.c.] \nonumber\\
-t_2\sum_{n,m}[a_{nm}^{\dag}(a_{nm+1}-a_{n+1,m})~~~~~~\nonumber\\
-b_{nm}^{\dag}(b_{nm+1}-b_{n+1,m}) + H.c.]~,~~~ 
\end{eqnarray}
where $E_a,E_b$ are site energies, and $t_1,t_2$  are  
the nearest-neighbor and next-nearest-neighbor hopping parameters.
Associated to $t_1$, the phase  $\gamma$ can be considered  as resulting 
from a periodic magnetic field (as in Haldane's picture \cite{Haldane}), 
which  does not generates any flux through the unit cell.

As usual, the Fourier transform 
$a_{nm}=\sum_{\vec{k}}a_{\vec{k}}~e^{i\vec{k}\vec{R}_{nm}}$ 
(and similarly for $b_{nm}$) helps in the Hamiltonian diagonalization.
With the choice of $\vec{a}_1$ and $\vec{a}_2$ as in Fig.3a, 
the Fourier transform of (1) reads:
\begin{eqnarray}
H=\sum_{\vec{k}}a_k^{\dag} a_k[E_a+2t_2\big(cos\vec{k}\vec{a}_1- 
cos\vec{k}\vec{a}_2\big)]  \nonumber \\
+\sum_{\vec{k}}b_k^{\dag} b_k[E_b-2t_2\big(cos\vec{k}\vec{a}_1- 
cos\vec{k}\vec{a}_2\big)]  \nonumber \\
-t_1\sum_k a_k^\dag b_k 
[e^{-i\gamma}(1+e^{-i\vec{k}(\vec{a}_1+\vec{a}_2)})~~~~~ \nonumber \\
+e^{i\gamma}(e^{-i\vec{k}\vec{a}_2}+e^{-i\vec{k}\vec{a}_1}) + H.c.].
\end{eqnarray}
By taking   $\{\vec{a}_1,\vec{a}_2\}$  along the  
Ox- and Oy-axis, respectively, and the zero energy such that $E_a+E_b=0$, 
with the notations:
\begin{eqnarray}
f(\vec{k})&=&cos k_x- cos k_y, \nonumber \\
g(\vec{k})&=& e^{-i\gamma}(1+e^{-i(k_x+k_y)})+
e^{i\gamma}(e^{-ik_x}+e^{-ik_y}) \nonumber \\ 
\Delta&=& (E_a -E_b)/2, 
\end{eqnarray}
the Hamiltonian (2) can be written in the  matrix form:
\begin{equation}
H=\sum_{\vec{k}}
    \left(\begin{array}{cc}
      a^{\dagger}_{\vec{k}}~ & b^{\dagger}_{\vec{k}} 
\end{array}\right)
        \left(\begin{array}{ccc}
                \Delta+2t_{2}f(\vec{k}) & -t_{1}~ g(\vec{k}) \\
                -t_1g^*(\vec{k}) & -\Delta-2t_2f(\vec{k})
        \end{array}\right)
    \left(\begin{array}{c}
            a_{\vec{k}} \\
            b_{\vec{k}}\\
\end{array}\right).
\end{equation}
The spectral properties and the symmetries of (4) can be conveniently analyzed 
by expressing the Hamiltonian $H(\vec{k})$ in terms of Pauli matrices:
\begin{equation}
H(\vec{k})=
-t_1 Re g(\vec{k},\gamma)~ \sigma_x 
+t_{1} Im g(\vec{k},\gamma) \sigma_y +
\big(\Delta+2t_{2}f(\vec{k}) \big)\sigma_z, 
\end{equation}
where the coefficients show the properties:  
\begin{eqnarray}
Reg(\vec{k},\gamma)&=&Reg(-\vec{k},-\gamma), \nonumber  \\
Img(\vec{k},\gamma)&=&-Img(-\vec{k},-\gamma), \nonumber\\
f(\vec{k})&=&f(-\vec{k}).
\end{eqnarray}
It is of interest to notice that the spectrum of any Hamiltonian 
of the form (5) is symmetric around the energy $\omega=0$,
exhibiting this {\it electron-hole  symmetry} 
no matter the  parameters $t_1, t_2, \gamma, \Delta$.
In order to prove the symmetry, it is sufficient to find an operator 
$\mathcal{P}$ that anticommutes with the Hamiltonian. The existence of  
such an operator ensures  that, if the eigenenergy $\omega_{\vec{k}}$ belongs 
to the spectrum, the same is true for $-\omega_{\vec{k}}$, 
i.e., the energy spectrum is symmetric. 
Indeed, one checks easily that the operator
\begin{equation}
\mathcal{P}=Im g(\vec{k})\sigma_x+Re g(\vec{k})\sigma_y
\end{equation}
anticommutes with (5). More than this, the  analytical expression of 
the eigenenergies can be obtained straightforward from (4):
\begin{equation}
\omega_{\pm}(\vec{k})=\pm\sqrt{\big(\Delta +2t_{2}f(\vec{k})\big)^2
+t_{1}^2~|g(\vec{k},\gamma)|^2} ,  
\end{equation}
showing explicitly a symmetric two-branch spectrum. 

A second peculiarity of the spectrum (8) is the {\it anisotropy},
which is manifest provided that $\Delta \ne 0$, and  is due to 
the obvious relation $f(k_x,k_y)=-f(k_y,k_x)$.

The {\it time-reversal} (TR) symmetry of the model is controlled
by the phase $\gamma$, associated to the nearest-neighbor hopping $t_1$. 
Indeed, the TR invariance condition $H(\vec{k})=H^*(-\vec{k})$ 
requires $g(\vec{k},\gamma)=g^*(-\vec{k},\gamma)$, while the function $g$ 
defined in (3) exhibits $g(\vec{k},\gamma)=g^*(-\vec{k},-\gamma)$,  
proving that the time-reversal invariance of the Hamiltonian 
is guaranteed only for $\gamma=0$.

In what concerns the {\it inversion} symmetry  $H(\vec{k})=H(-\vec{k})$,
it is easy to see from (8) that the symmetry is fulfilled only if 
$g(\vec{k},\gamma)=0$. 

The energy spectrum (8) describes either a {\it semimetal} or an 
{\it insulator}, depending on the considered point in the parameter 
space $\{\gamma, \Delta, t_2\}$, when $t_{1}$ is taken  for energy 
unit ($t_{1}=1$).

Let us  discuss first the occurrence of the semimetallic phase.  
The touching  of the  bands at $\omega=0$ occurs provided that the two
terms under the square root in (8) vanish {\it simultaneously}. 
It is to distinguish three cases:

a) If $\gamma=0$, the function $g(\vec{k},\gamma=0)$ vanishes along 
two lines $k_x=\pi$ and $k_y=\pi$ in the BZ. Along the first line one has 
$f(\pi,k_y)=-2 cos^{2}k_y/2$, resulting the dispersion
\begin{equation}
\omega_{\pm}(\pi,k_y)=\pm(\Delta-4t_2cos^2k_y/2).
\end{equation}
Then, as long as  $0< \Delta/4t_2 < 1$, 
the condition  $\omega_{\pm}(\pi,k_y)=0$ 
is satisfied at $k_y=\pm 2 acos\sqrt{\Delta/4t_2}$,  meaning that
there are two points belonging to  BZ where the bands touch each other, 
giving rise to the semimetallic phase.
In the particular case  $\Delta/4t_2 = 1$, the expression (9) vanishes
at $k_y=0$, meaning the existence of  a single touching point 
in the BZ at $\{k_x=\pi,k_y=0\}$. 
The same is obtained  for $\Delta=0$, the touching point
being now  $\{k_x=\pi,k_y=\pi\}$.

A similar discussion  carried out for the other line ($k_y=\pi$)
indicates that, as long as $\Delta$ and $t_2$ are taken  positive,  
the condition $\omega=0$ cannot be fulfilled along it.

b) If  $\gamma \ne 0$, the function $g(\vec{k},\gamma)$ vanishes only  
at four points located at the edges of BZ, namely $\vec{k}_{1,2}=(0,\pm\pi), 
\vec{k}_{3,4}=(\pm\pi,0)$. 
If we assume now  $\Delta=t_{2}=0$, the band touching occurs at the  four 
mentioned points, around which the energy dispersion is linear (see Fig.2), 
meaning in fact {\it two} cones per BZ.

c) If  $\gamma \ne 0$ and  $\Delta/4t_2=1 $, the band touching occurs only  at
the  two points $\vec{k}_3=(\pi,0)$, $\vec{k}_4=(-\pi,0)$,
where $f(\vec{k})=-2$, resulting a semimetal with a {\it single} cone per BZ.
(As observed  also by Hou \cite{Hou}, the cone position in BZ 
depends on the sign of $\Delta$. 
So, if $\Delta/4t_2=-1 $, the band degeneracy occurs now at  
$\vec{k}_1,\vec{k}_2$, 
the spectrum being rotated with  $\pi/2$ compared to the other case.)

Except for the above mentioned cases the energy spectrum is gapped. 
However, it is to find out under what circumstances the insulating phase 
is a topological one.  
A specific feature of the model is that  the time-reversal symmetry breaking 
is  necessary, but not  sufficient, for the appearance of the topological 
insulator phase. The reason is that the phase $\gamma$ breaks indeed the 
TR symmetry, but cannot open the gap.  
The statement in the case c) above says that the existence of the gap requires 
$\Delta\ne 4t_2$, meaning  that either $\Delta$ or $t_2$ must be 
different from zero.

Based on the calculation of the Chern number, a topological phase showing 
anomalous quantum Hall effect is predicted in \cite{Hou} under the condition  
$\Delta/4t_2 < 1$, which implies that $t_2$ should be the non-vanishing parameter. 
In the opposite case ($\Delta/4t_2 > 1$), the system behaves as a conventional 
bulk insulator. 
In the next sections, this problem  will be analyzed in terms of the spectral 
properties of confined systems, by looking for topologically protected  
edge states in the ribbon (strip) and plaquette geometry. 
%Fig2
\begin{figure}
\includegraphics[angle=-00,width=0.30\textwidth]{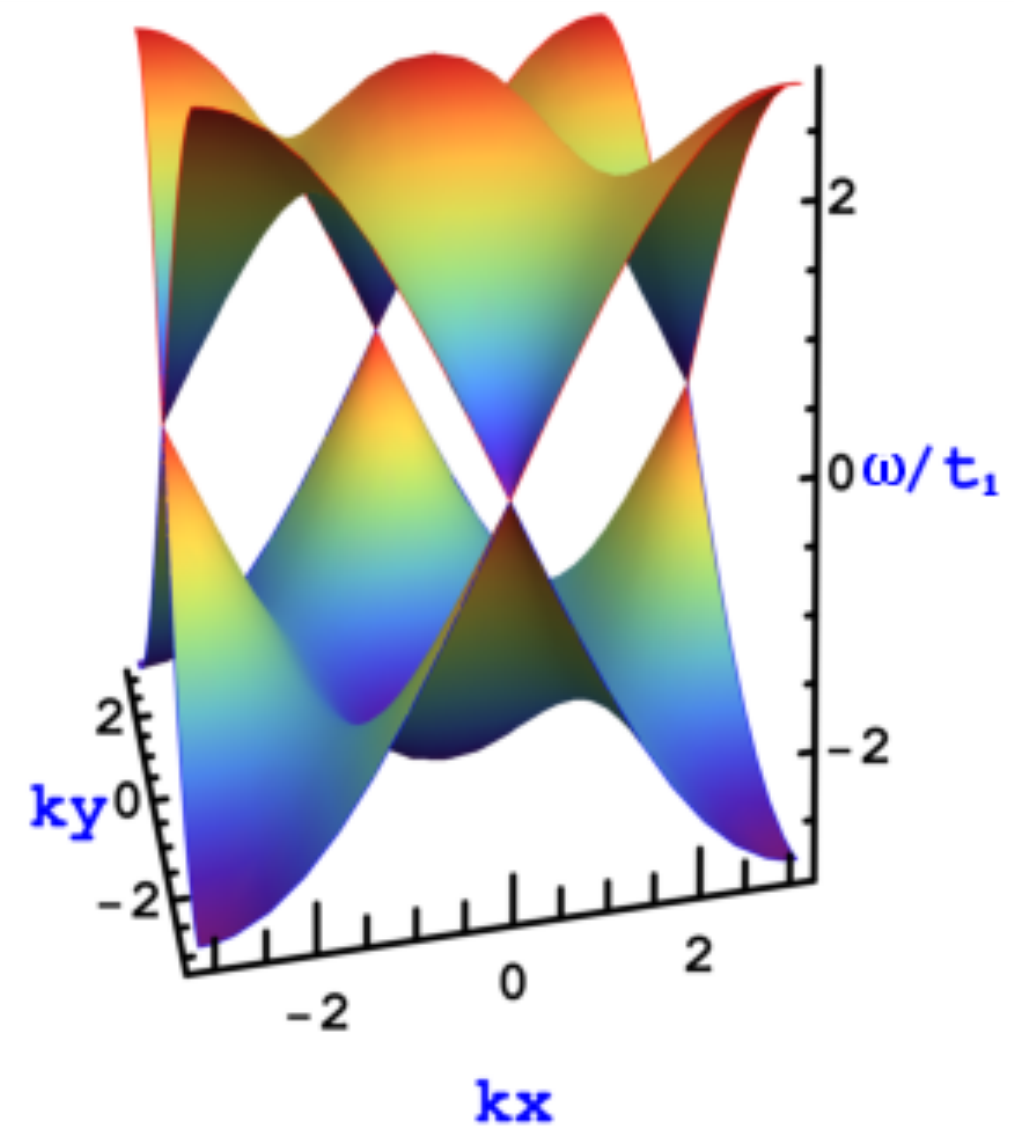}
\caption{(Color online) The semimetallic case b) in the text showing 
band touching 
for $t_{2}=\Delta=0, \gamma=\pi/4$. } 
\end{figure}
\section{edge states in  the  ribbon geometry}
We study  two kinds of ribbons obtained  by cutting the  edges  
in different manners: parallel to  the A-A and A-B direction, respectively, 
the angle between the two directions being $\pi/4$.
The intention is to detect the differences, if any, in the spectral properties 
coming from the different atomic connectivities at the ribbon edges 
in the two cases.
Since the ribbon cut along the A-A chain ends with a B-B chain,
this type of ribbon will be called in what follows as the AA/BB ribbon.
The other ribbon gets the both cuts along the same type of chain, and
it will be called the AB/AB ribbon.

Technically, one has to chose the unit cell in two different ways 
corresponding to the two types of ribbons (see Fig.3). Next,
one performs the Fourier transform of the Hamiltonian along the direction 
parallel to edges, and then, the ribbon energy spectrum is calculated by   
numerical diagonalization.
Still, the  existence of the flat edge states
shown by the AA/BB ribbon in Fig.4a is proved analytically.

The presence of topologically protected or dissipative edge states in
a gapped spectrum is a rather common feature of confined 2D systems. 
However, in this section, we prove the  existence  of {\it dispersive} 
current-carrying  edge states 
in the {\it gapless} phase of the ribbon, which seems to be a new 
conceptual aspect in the frame of topological states.

%Fig3
\begin{figure}
\hskip-2cm
\vskip1cm
\includegraphics[angle=-00,width=0.5\textwidth]{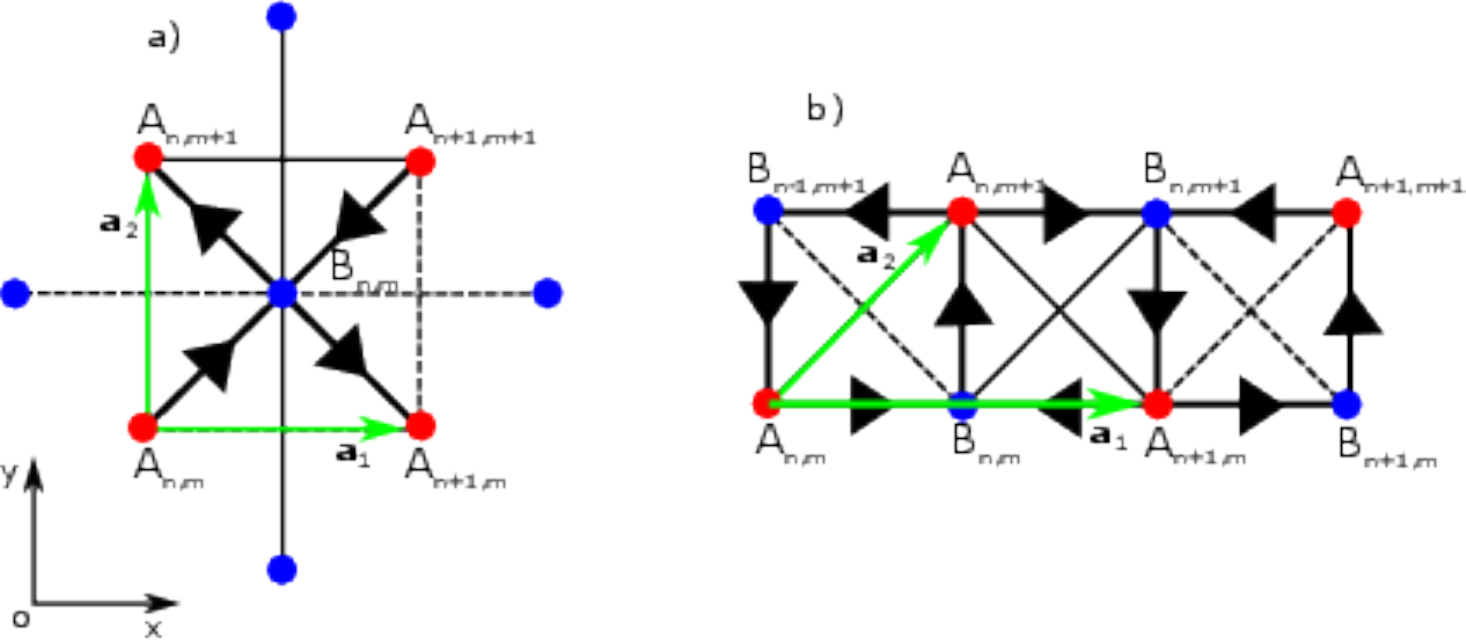}
\caption{(Color online) The unit cells defined by the vectors $\vec{a}_1$ 
and $\vec{a}_2$
(in green) for the ribbon cut parallel to A-A atomic chain in (a), 
and parallel to A-B chain in (b). }
\end{figure}

The discussion of the spectral properties in the ribbon geometry 
will be carried out in several steps. First, we confirm the 
semimetal phase, and next we consider the insulating phases. 
The rather many possible cases are summarized at the end 
of the section in the phase diagram Fig.8.

It turns out that the semimetal  exists under the  same constraints
as for the infinite system:\\
1. in the case $\gamma\ne 0$ and $\Delta=t_2=0$ (similar to the case b)
in the discussion of the infinite lattice),  
the semimetallic character is revealed by the both types of ribbons,  
although there are non-trivial differences between them, obvious in Fig.4. 
For instance, the flat band at $E=0$ in the spectrum of the AA/BB-ribbon 
shows a striking similarity to  the case of the well-known zig-zag graphene 
ribbon \cite{Wakabayashi}. 
We show in what follows that the flat band is composed of edge states, 
which occur at any momentum $ k \in [0,\pi]$ no matter the value of 
the phase $\gamma$. To this end, we perform the Fourier transform of 
the Hamiltonian (1) along the ox-axis (parallel to AA-edge) and  impose 
vanishing  boundary conditions on the perpendicular direction. 
The resulting Hamiltonian for the AA/BB-ribbon reads:
\begin{eqnarray}
H(k)=\sum_{m=1}^M (\Delta +2t_2cosk) [a_{km}^{\dag}a_{km} -
b_{km}^{\dag}b_{km}] \nonumber\\
-t_2\sum_{m=1}^M [a_{km}^{\dag}a_{km+1} -b_{km}^{\dag}b_{km+1}
 +H.c.] \nonumber\\
-t_1\sum_{m=1}^M [a_{km}^{\dag}b_{km} (e^{-i\gamma}+e^{i\gamma-ik}) +
~~~~~~\nonumber \\
a_{km}^{\dag}b_{km-1} (e^{i\gamma}+e^{-i\gamma-ik}) + H.c.]~,
\end{eqnarray}
where the index $m$ counts the unit cells along the ribbon width.
(The AB/AB-ribbon can be addressed similarly.)

% Fig4
\begin{figure}
        \includegraphics[angle=-0,width=0.23\textwidth]{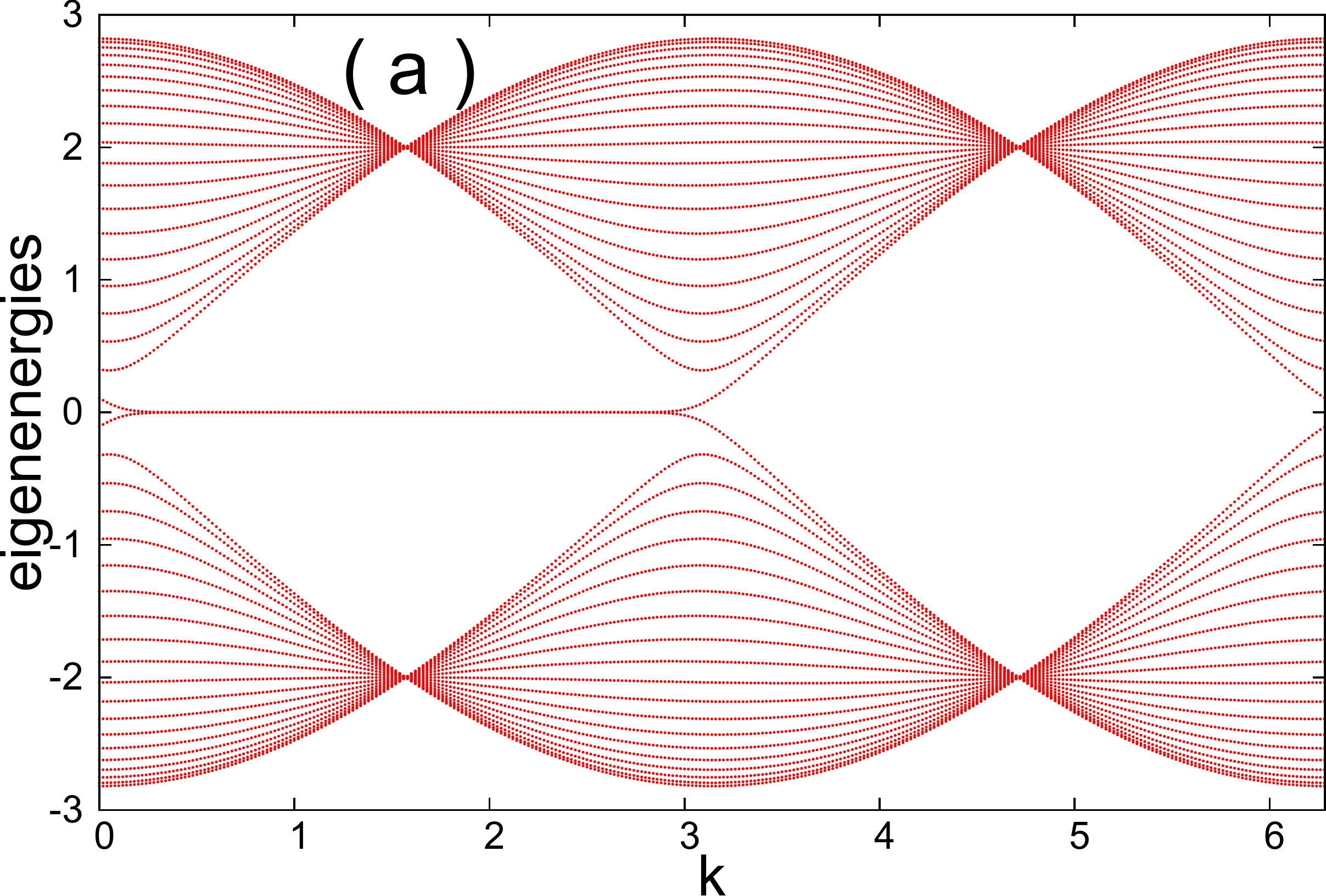}
        \includegraphics[angle=-0,width=0.23\textwidth]{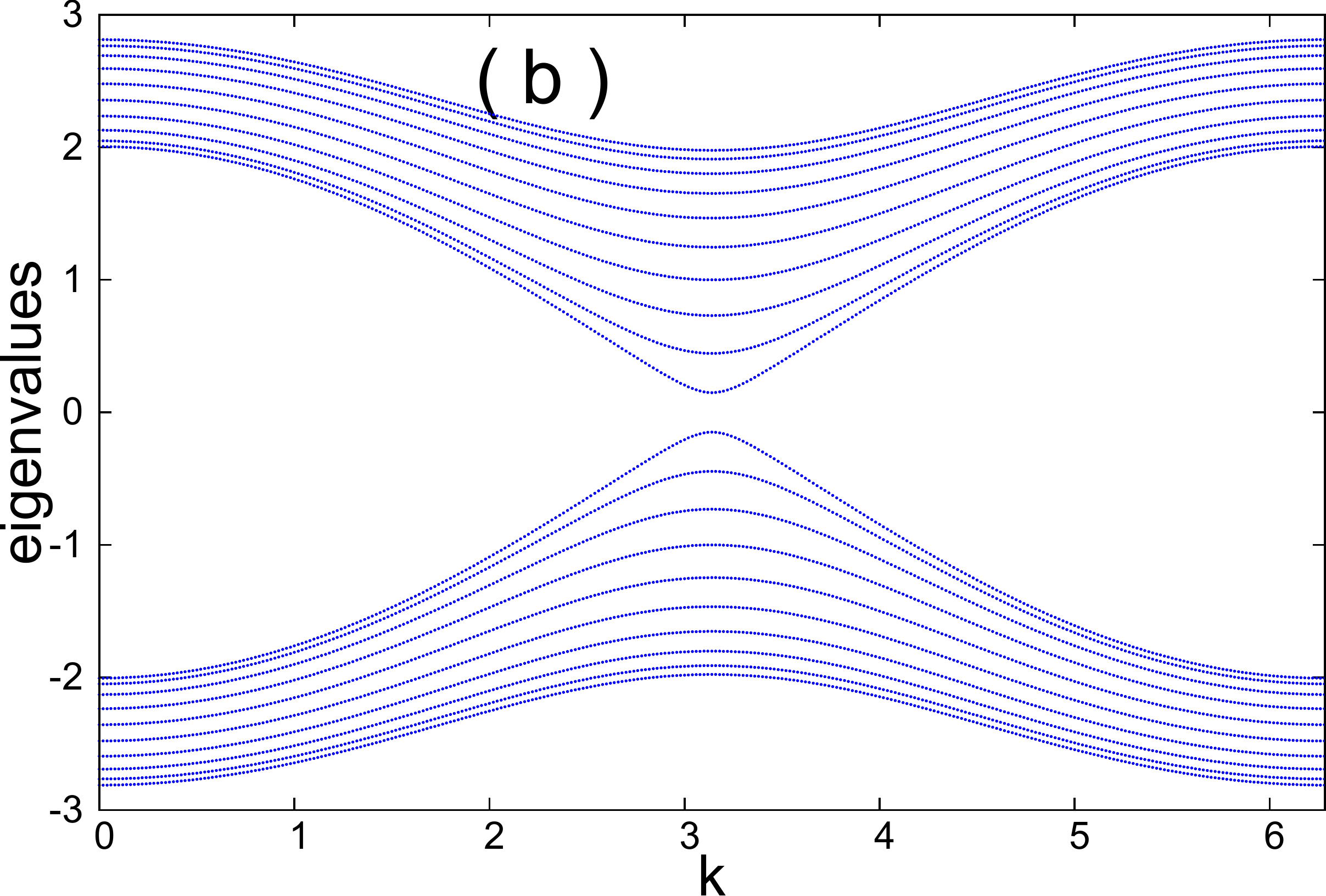}
\caption{(Color online) Semimetallic spectrum under the conditions 
$t_2=\Delta=0$, $\gamma=-\pi/4$ for the AA/BB (in red) and AB/AB 
(in blue) ribbons. The width is $M=20$ unit cells.}
\end{figure}

Looking for  the eigenfunctions of (10) as
\begin{equation}
|\Psi(k)\rangle=\sum_m (\alpha_{km}a_{km}^{\dag}+
\beta_{km}b_{km}^{\dag})|0\rangle,
\end{equation}
in the case under discussion  ($\Delta=t_2=0$), the coefficients satisfy 
the following linear equations ($t_1=1$):
\begin{equation}
-(e^{i\gamma}+e^{-i\gamma+ik})\alpha_{km}+ (e^{-i\gamma}+
e^{i\gamma+ik})\alpha_{km+1}=E(k)\beta_{km}\nonumber\\
\end{equation}
\vskip-0.72cm
\begin{equation}
-(e^{-i\gamma}+e^{i\gamma-ik})\beta_{km}+ (e^{i\gamma}+
e^{-i\gamma-ik})\beta_{km-1}=E(k)\alpha_{km}.
\end{equation}
We are interested in the behavior of the wave function at $E=0$, 
where  Eq.(12) gives
\begin{equation}
\alpha_{k,m}=x^{m-1} \alpha_{k,1},~~ 
 x=\frac{e^{i\gamma}+e^{-i\gamma+ik}}
{e^{-i\gamma}+e^{i\gamma+ik}}~.
\end{equation}
Under the condition $|x|<1$, the coefficient $\alpha_{km}$ shows 
the maximum value at the edge  m=1 and decays  monotonically 
versus the other edge m=M. (It is trivial to show that $\beta_{km}$ 
behaves oppositely.) One concludes that the wave function 
$|\Psi(k)\rangle=\sum_m \alpha_{km}a_{km}^{\dag}|0\rangle$,  
corresponding to the eigenvalue $E(k)=0$, describes an edge state 
localized near the edge $m=1$. 
Since $|x|^2=(1+cos(2\gamma -k)/(1+cos(2\gamma +k))$, 
the condition $|x|<1$ can be easily rewritten as $sin(2\gamma) sink <0$ 
and  is satisfied for any  $k$ in the range $[0,\pi]$, if  $\gamma$ 
is assumed  to be negative. This result confirms the numerical data 
shown in Fig.4a.

One has to comment for the sake of accuracy that, as it is known for 
such kind of problems \cite{Wakabayashi}, the perfect degeneracy 
at  $E=0$ occurs only in the limit of infinite wide ribbon 
($M \rightarrow \infty$), otherwise the spectrum remains quasi-degenerate.
The reason for such a behavior is  the superposition of the edge states 
localized near opposite edges, which  although small is still non-zero 
at finite width \cite{OA}.

2. The  semimetallic character is preserved along the line $\Delta/4t_2= 1$
no matter  whether the TR symmetry is broken ($\gamma\ne 0$) or not 
($\gamma=0$). The situation is similar to the cases a) and c) discussed 
previously for the infinite lattice. Again, this is true for  both ribbons, 
existing however qualitative differences illustrated in Fig.5.
%Fig5
\begin{figure}
\hskip-1cm
\includegraphics[angle=-0,width=0.23\textwidth]{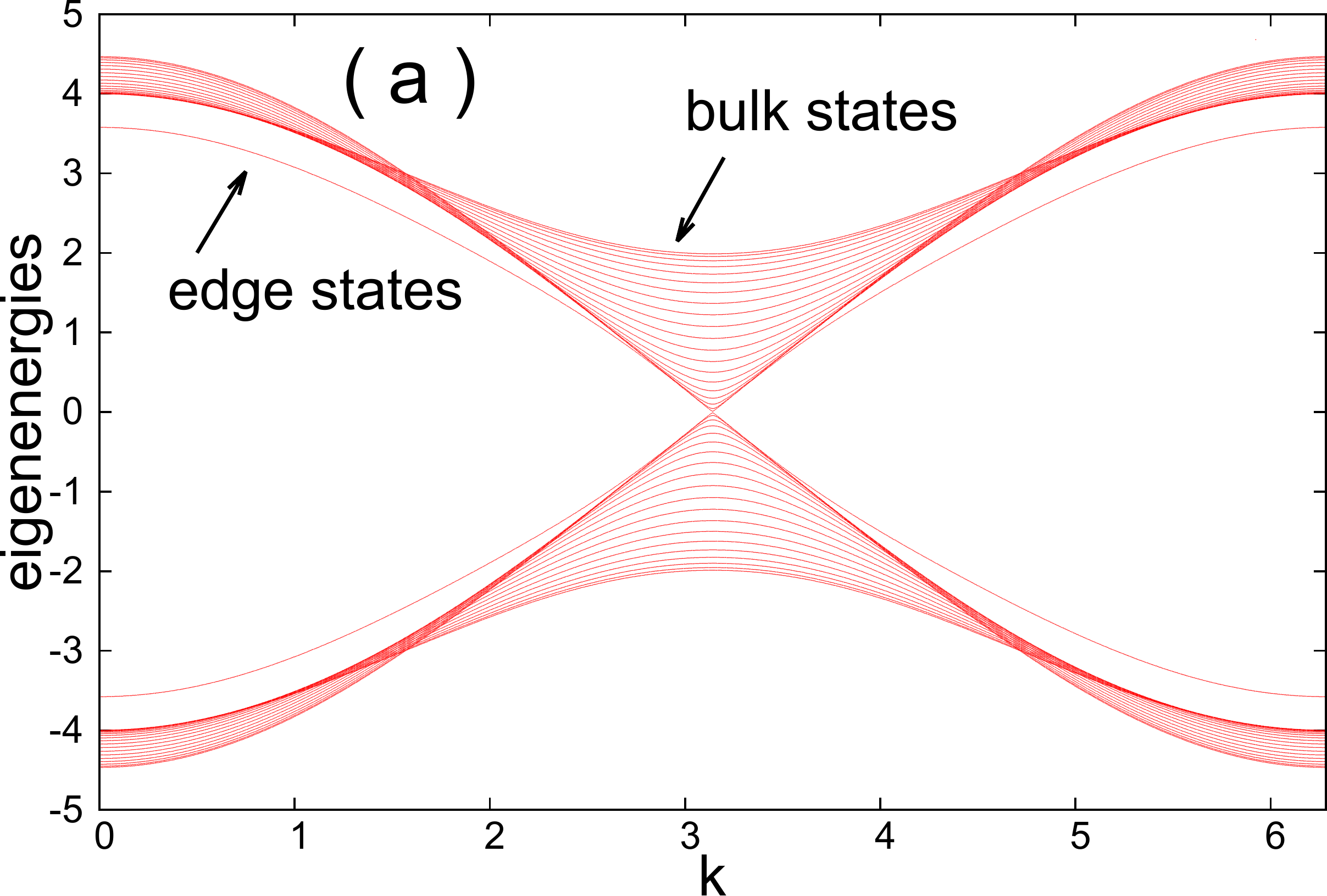}
\includegraphics[angle=-0,width=0.23\textwidth]{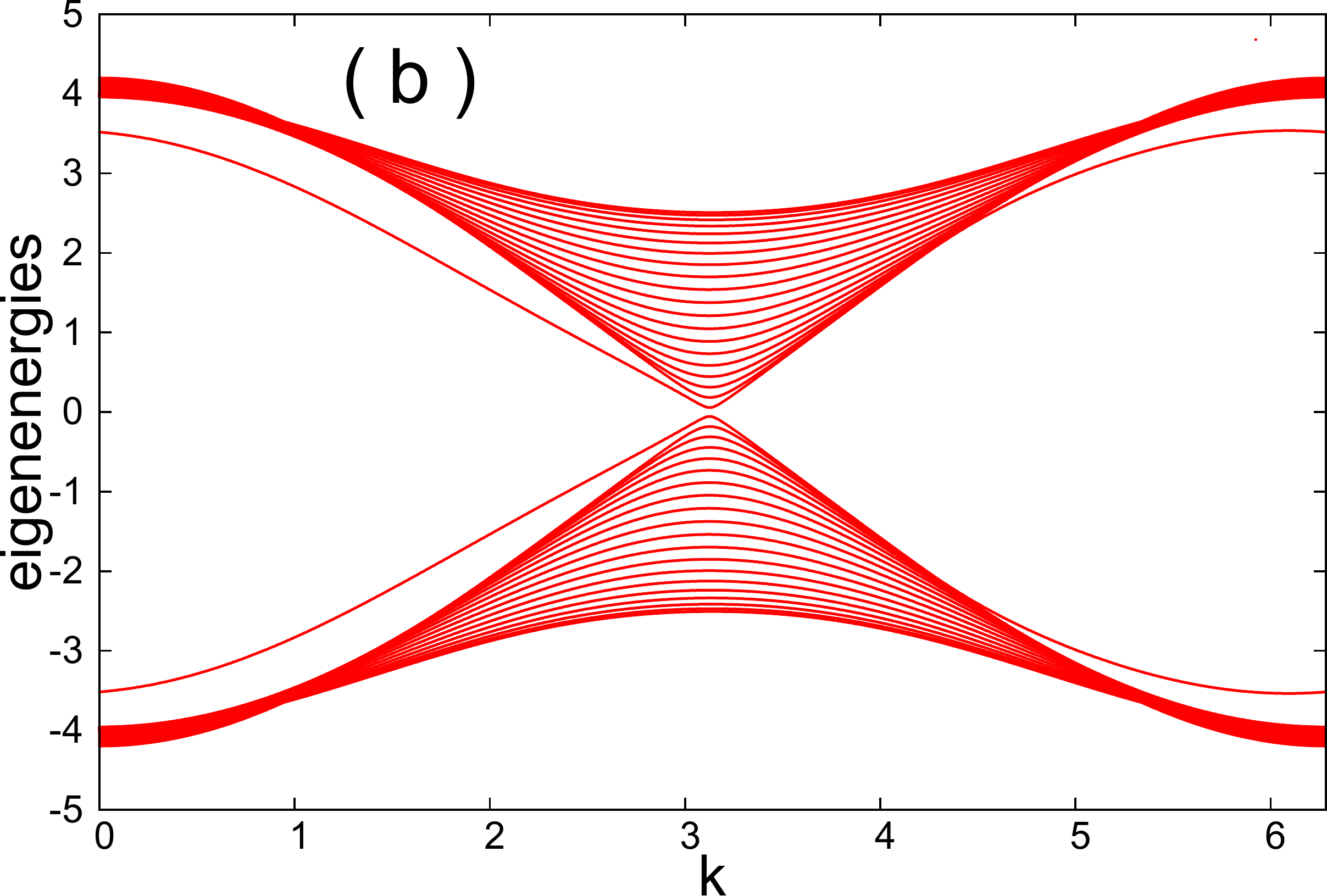}
\includegraphics[angle=-0,width=0.23\textwidth]{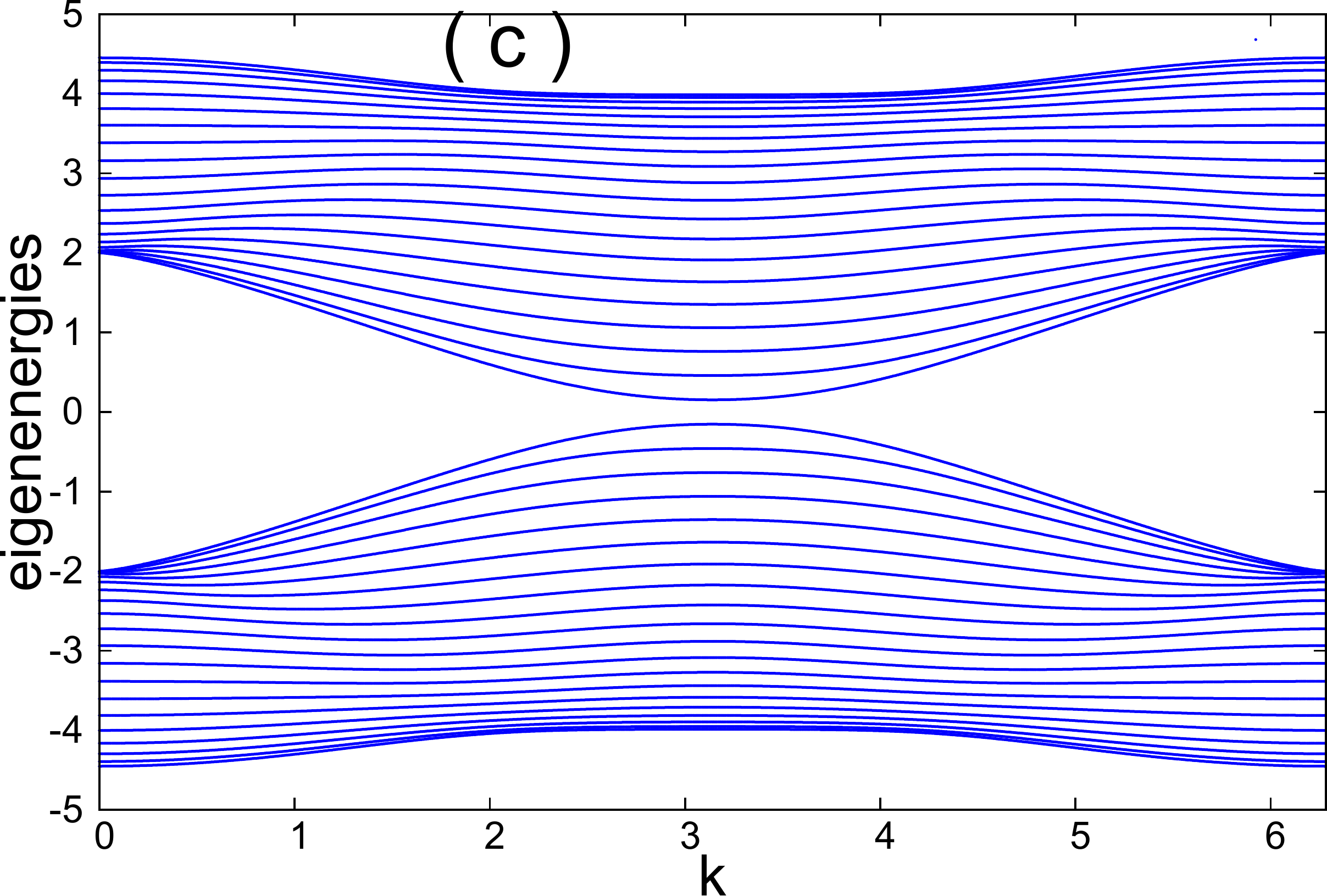}
\caption{(Color online) Semimetallic spectrum under the conditions  
$t_2=0.5, \Delta=2$.  Branches consisting of edge states  are 
visible for the AA/BB ribbon (in red).  The branches are symmetric
in (a) at $\gamma=0$, and become asymmetric in (b) at  non-vanishing
phase ($\gamma=\pi/8$). The  AB/AB ribbon  
does  not exhibit such features (the case $\gamma=0$ is shown in (c)).}
\end{figure}

The important finding in this case is that the AA/BB-ribbon exhibits 
dispersive edge modes as shown in Fig.5a and Fig.5b. Since the spectrum 
is gappless, inevitably, these  modes are  energetically embedded in 
the continuum of the bulk states.
So, we notice in the panels (a) and (b) of Fig.5 that some eigenvalues 
are organized in branches which are detached from the rest of the spectrum. 
This feature is missing for the AB/AB ribbon (Fig.5c).

The geometrical localization of the corresponding wave functions 
can be inspected by calculating  $|<i|\Psi>|^2$, meaning
the projection of the  wave function $|\Psi>$ on the lattice 
sites $i$ along the width of the ribbon. Fig.6 shows that such a state
(shown in blue) is peaked at the ribbon edge,
while a  state in the bulk (shown in red) covers the middle of the ribbon.

Another significant observation is the symmetry  of the edge state 
branches around $k=\pi$ at $\gamma=0$ in Fig.5a. The meaning is that, 
at any energy $E$, there are two  states of opposite velocity $dE/dk$ at 
the same edge, resulting a vanishing total current carried along edges.
On the other hand, if  $\gamma \ne 0$ as in Fig.5b, the edge 
state branches  become asymmetric,
and  non-vanishing currents, running in opposite directions along 
the two edges, are allowed. In other words, the AA/BB-ribbon  with 
n.n.n. hopping and broken TR symmetry, supports chiral edge states
even in the semimetallic phase. 
The fate of these  states in the case of finite size  plaquettes will 
be examined in Sec.IV.

%Fig6
\begin{figure}
\includegraphics[angle=-0,width=0.28\textwidth]{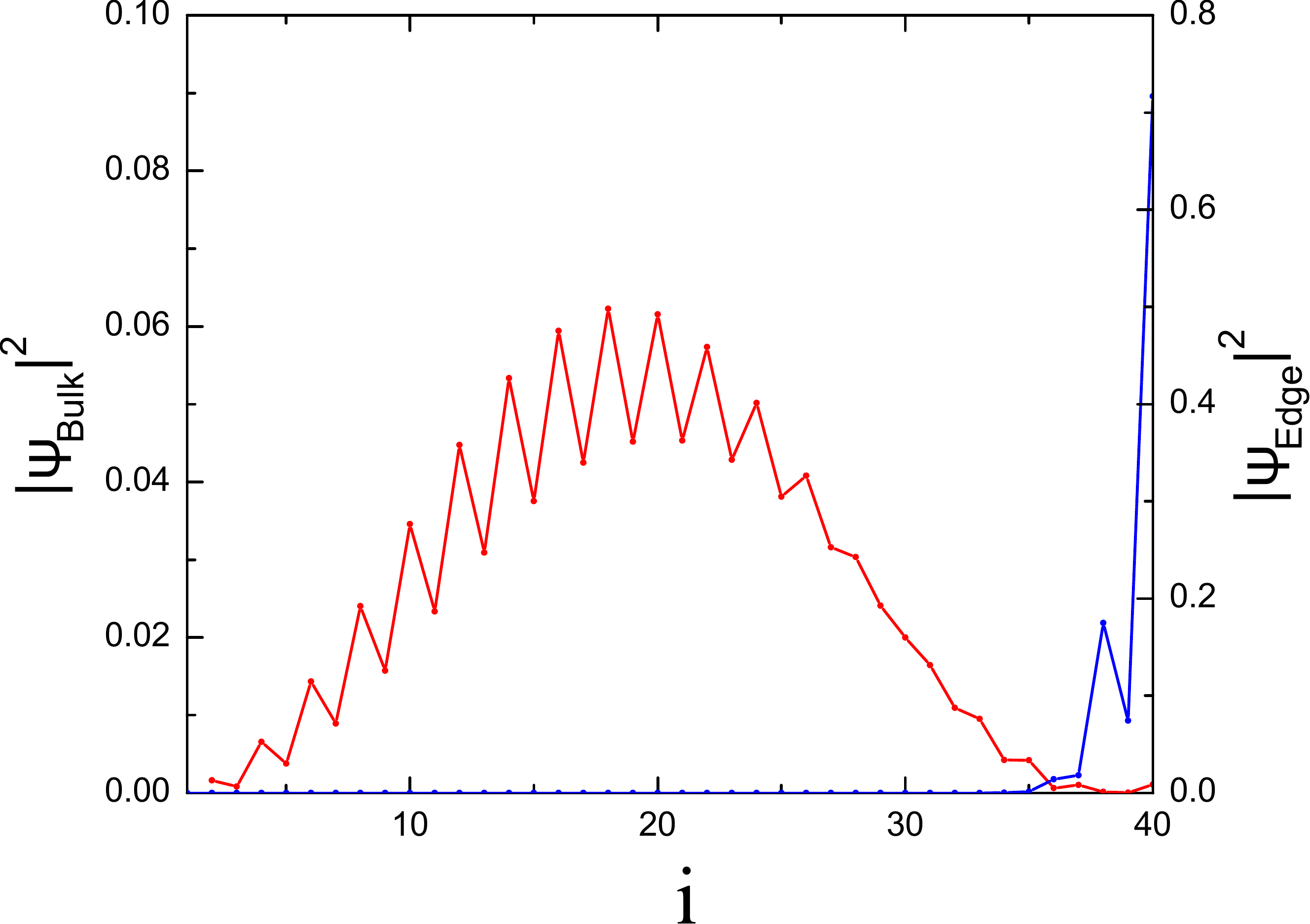}
\vskip-0.1cm
\caption{(Color online) The local distribution of two wave functions
at $E\simeq -1$ in the semimetallic phase for the AA/BB ribbon 
with M=20 cells:  the  edge state (blue) at  $k=0.72\times\pi$ and the
 bulk state (red) at  $k=0.82\times\pi$. The index 'i' counts the
sites along the ribbon width (other parameters:  $t_2=0.5, \Delta=2, 
\gamma=\pi/8$).}
\end{figure}

3. Finally, we look  for the Chern insulating phase in the ribbon geometry,
distinguished  by the presence of chiral edge states in a {\it gapped} spectrum.  
The spectra in Fig.7 demonstrate that the breaking of the TR symmetry  
(i.e., $\gamma\ne 0$) is not sufficient for the occurrence of the Chern insulator. 
Indeed, there is a separation line  $\Delta/4t_2=1$  above which 
the system  is a conventional bulk insulator (as in the panels (c) and (d)), 
while below it (for $\Delta/4t_2<1$) the system becomes a topological
insulator. 
The result indicates that the existence  of the edge states in the gap is 
dictated by two conditions: $\gamma \ne 0$ {\it and} $t_2 \ne 0$. 
This behavior is the same for the both types of ribbon, and 
one has to note that the separation line is that one along which the system 
becomes a semimetal.
%Fig7
\begin{figure}
        \includegraphics[angle=-0,width=0.23\textwidth]{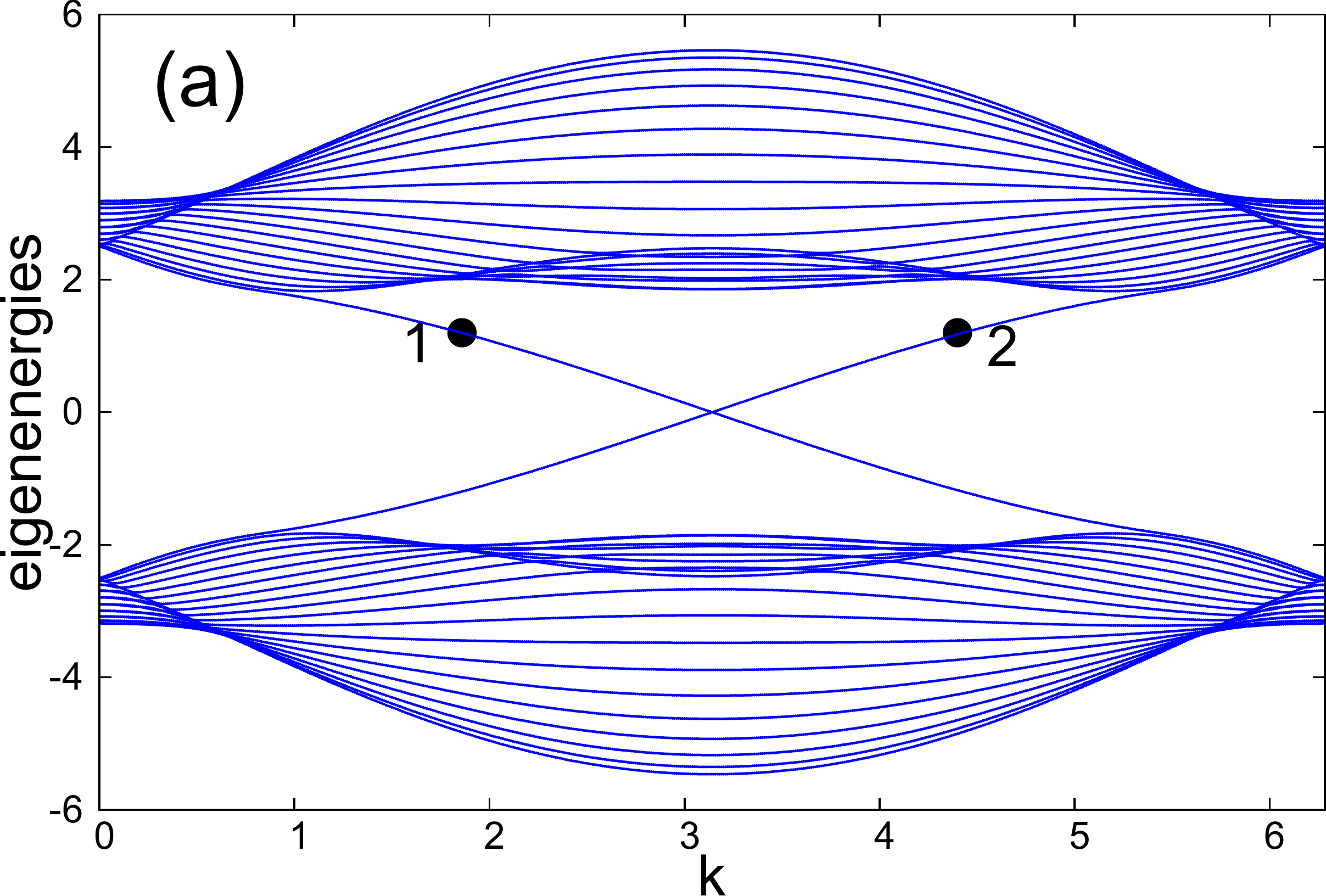}
        \includegraphics[angle=-0,width=0.23\textwidth]{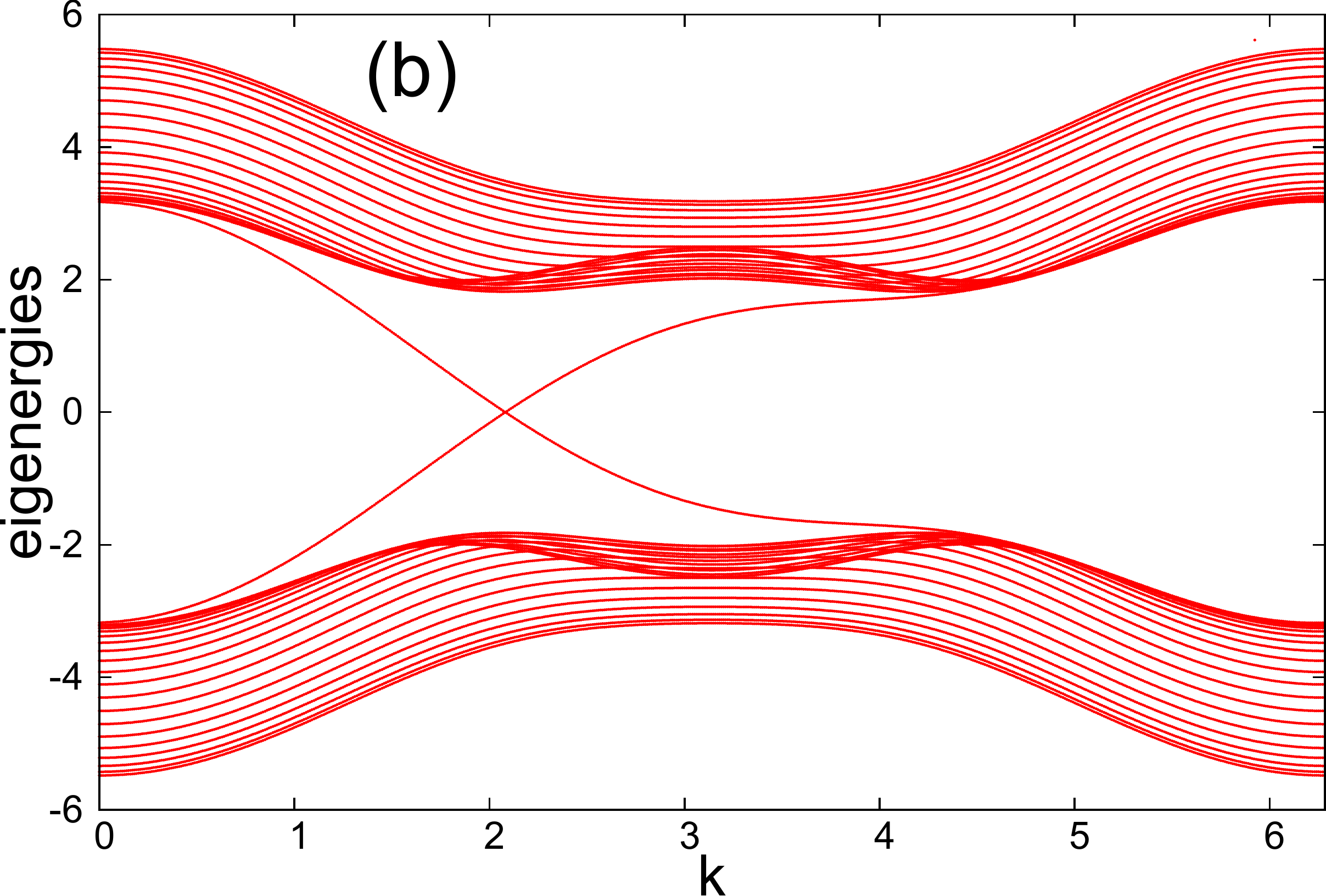} \\
        \includegraphics[angle=-0,width=0.23\textwidth]{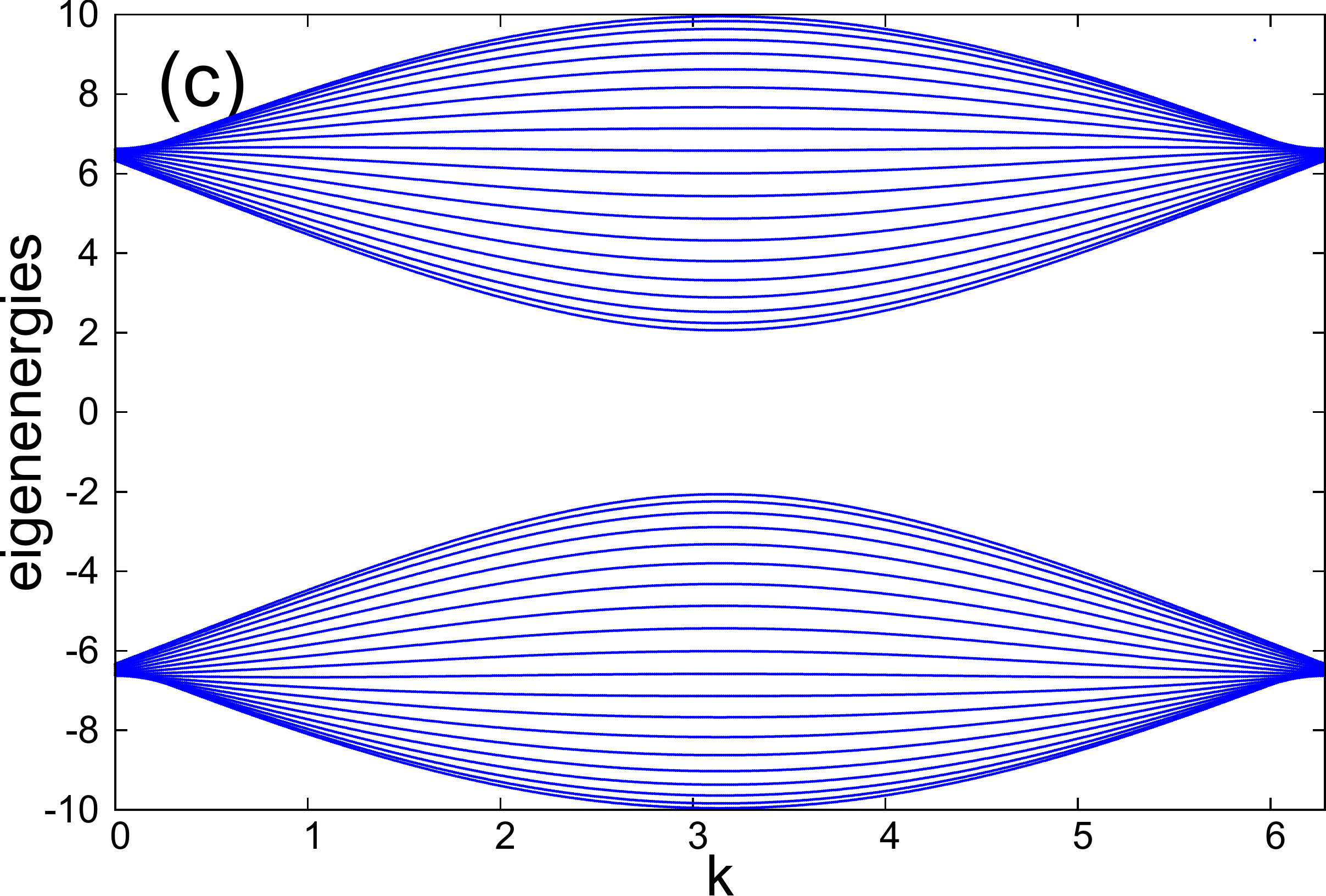}  
        \includegraphics[angle=-0,width=0.23\textwidth]{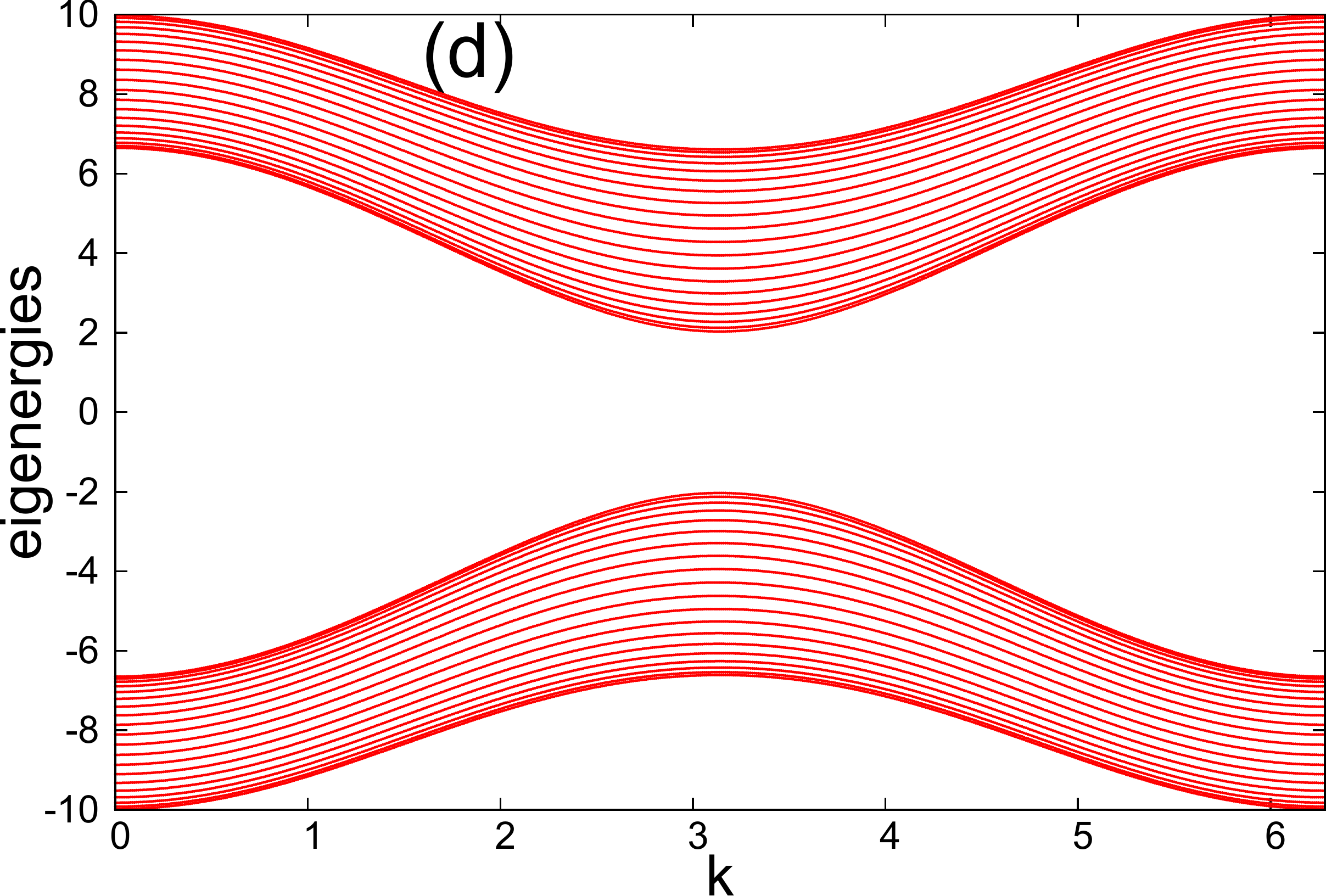} 
\caption{(Color online) (a), (b): The ribbon energy spectrum as 
function of the momentum $k$ in the 
Chern insulating phase showing edge states in the gap 
($\gamma=\pi/4,t_2=1,\Delta=1.5$). (c), (d): The spectrum in the 
conventional insulating phase ($\gamma=\pi/4,t_2=1,\Delta=6$).
The spectra for  AB/AB ribbon are presented in blue, those for AA/BB ribbon 
are in red) }
\end{figure}

However, one has still to decide  whether the states are protected or not 
against the backscattering. To this end, we observe  that the edge states 
indicated by "1" and "2" in Fig.7a are running at the same energy in opposite 
directions (i.e., show different chirality), but  prove to be located at 
{\it opposite} sides of the ribbon. Then, one concludes that they are robust against  
elastic processes and cannot undergo backscattering.

The results concerning the spectral properties are collected 
in the phase diagram Fig.8. 
%Fig8
\begin{figure}
\includegraphics[angle=-0.0,width=0.4\textwidth]{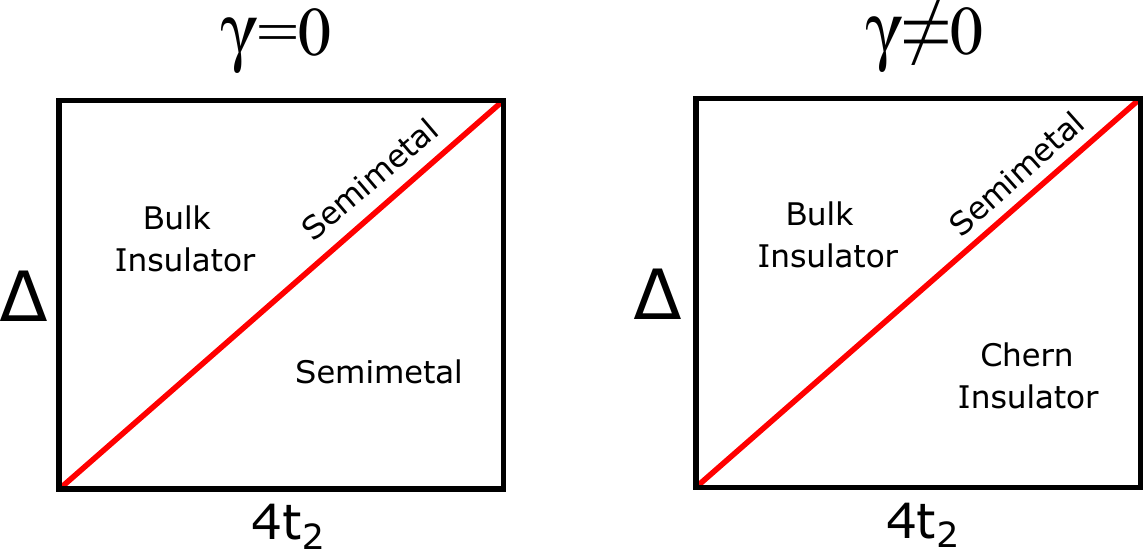}
\caption{The phase diagram in the space of relevant 
parameters \{$\Delta,t_2$\}, with and without time-reversal symmetry 
(left and right panel, respectively).}
\end{figure}
\section{plaquette geometry:~ edge states and anomalous quantum Hall effect} 
The attention will be focused now on the transport properties  and 
disorder effects, the study being carried out in  finite size 
(plaquette) geometry. One starts with the spectral properties of 
the plaquette, and  continues with the calculation of the Hall effect for  
clean and disordered systems.

1. {\it Spectral properties.}    
In the previous section, we identified edge states 
in the  ribbon geometry, when vanishing boundary conditions 
are applied  along one direction only. Our  goal now is to examine 
the  edge states in the more physical situation  of confined systems 
obtained by imposing vanishing  conditions all around the perimeter.  
Actually, such a  plaquette is obtained by cutting the ribbon perpendicularly,  
but taking care to ensure a  vanishing  flux through the plaquette area.
One gets two types of plaquettes: with AB and  AA (or BB) atomic  sequence 
along the  perimeter, respectively.

We expect to get  edge states running along the boundary
at least in some range of parameters. One may  also anticipate  that, 
in case of broken TR symmetry, the edge states  would get chirality 
and support {\it anomalous} quantum Hall effect, detectable in a  
four-lead Hall device.  
This picture corresponding to a description in terms of edge states 
of the  Chern-type topological insulator is developed in what follows.

In what concerns the spectral properties of the finite system,
we  start underlying  the relevance of the case $\Delta/4t_2= 1$,
which  appeared to be important in the previously  discussed
geometries. 
Fig.9a, which shows the dependence of the spectrum on $\Delta/4t_2$ 
at $\gamma \ne 0$, is  illustrative in this respect:  one notices that
the gap is filled with (edge) states  if  $\Delta/4t_2< 1$ and it is clean
in the opposite case. This demonstrates the transition from the
Chern topological insulator to the  conventional insulator, crossing the
situation where the gap closing  indicates a semimetal.
The chirality and robustness of the edge states will be discussed below.

The dependence of the energy spectrum on the  phase $\gamma$   
is depicted in Fig.9b, in the regime $\Delta/4t_2 <1$.
This description is the analog of the Hofstadter-type energy 
spectrum of the {\it confined} 2D electron gas, expressed as function 
of the external magnetic flux. Because of this analogy, 
we shall equivalently call the phase $\gamma$ as a  'local-flux'. 
However, instead of the butterfly picture created by the sequence of 
bands and gaps in the Hofstadter spectrum,  the spectrum in Fig.9b
contains a single gap and exhibits a 'binocular' aspect. 

The eigenvalues as function of $\gamma$ are obtained by the  numerical 
diagonalization of the Hamiltonian matrix. Besides the 
electron-hole symmetry, some other properties are notable in Fig.9b:
 i) the spectrum is periodic with the period $\delta\gamma=\pi$, 
ii)  the spectrum shows mirror symmetry around $\gamma=\pi/2$, 
iii) the gap vanishes at $\gamma=0,\pi/2$ where the  system becomes 
semimetallic,
iv) the gap width depends on the 'local-flux' and reaches its  
maximum at $\gamma=\pi/4$ and $3\pi/4$. 

The emergence of chiral edge states in the gap  is the major
property of the energy spectrum of the finite plaquette in the 
case $0<\Delta/4t_2 <1$.
The edge character can be shown as in the ribbon case
by  calculating numerically the electron distribution $|<i|\Psi>|^2$ 
for any state $|\Psi>$ in the gap, where $i$ stands for any lattice site of
the plaquette. It turns out that, indeed, the gap states 
are located near the  boundaries, such a state being depicted in Fig.10.
%Fig9
\begin{figure}
	\includegraphics[angle=-0,width=0.5\textwidth]{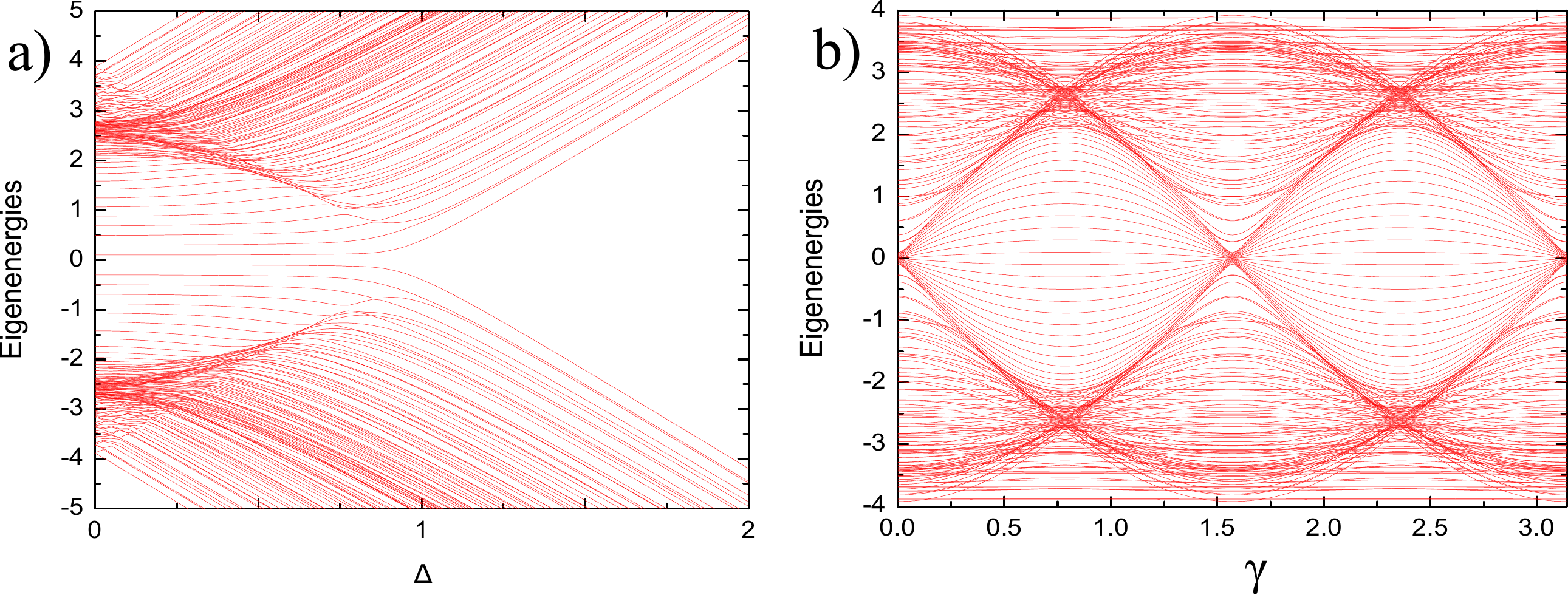}
\caption{(a) The energy spectrum of AB plaquette as function of the 
staggered energy $\Delta$ (in units $4t_2$) at $\gamma=\pi/4$.
(b) The energy spectrum as function of the phase $\gamma$ at $\Delta=0, t_2=1.$}
\end{figure}
%Fig10
\begin{figure}
        \includegraphics[angle=-0,width=0.28\textwidth]{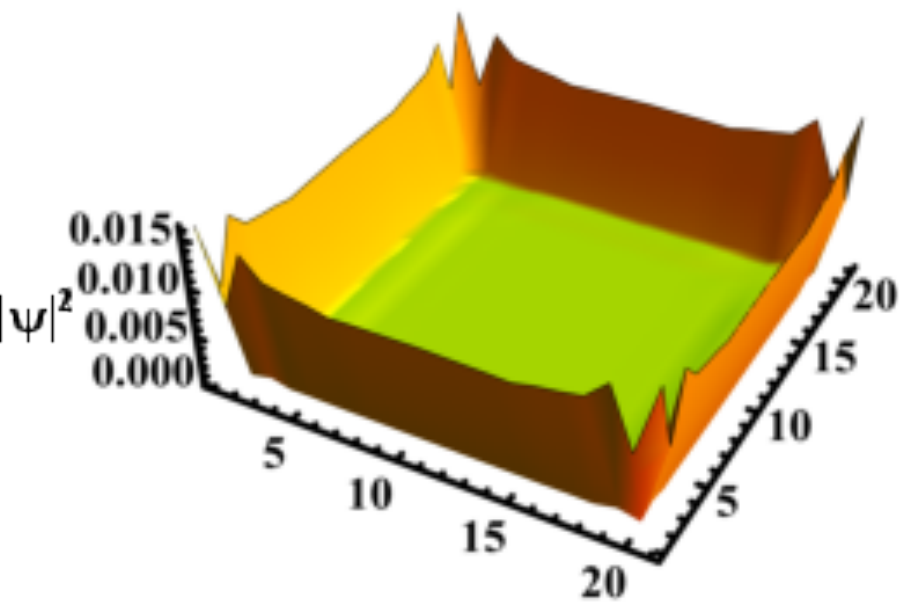}
\caption{(Color online) Electron density $|\Psi(r)|^2$ corresponding to
the  edge state at  $E=-1.21379$ for an AB plaquette of dimensions 
21$\times$20 cells 
($\gamma=\pi/4$, $\Delta=0$).}
\end{figure}

The confirmation of the chirality, meaning that the edge states  carry the 
electron current in a  given (clock/anticlockwise) direction, 
can be  achieved  by showing the {\it one-way} behavior of the 
transmission coefficients in the four-lead Hall device, 
which is the suitable arrangement for investigating the transport properties.

2. {\it Transport properties}.
The goal  is now to calculate the transverse (Hall) conductance presumed
to be non-zero by reason of time-reversal breaking induced by the 
local flux $\gamma$. On the other hand, one has to keep in mind that the 
total flux through the plaquette vanishes. Then,  an eventual integer value 
of the Hall conductance obtained by the calculation would confirm the 
anomalous quantum Hall effect carried by chiral edge states located  
in the gap of a  Chern insulator.

In order to do that, we simulate a Hall device by attaching four leads 
to the plaquette and  apply afterwards the Landauer-B\"uttiker formalism. 
The role of the leads is to inject current in the system and to 
enable the measuring of voltages.The approach requires the knowledge 
of the full Hamiltonian that describes the finite plaquette, 
the leads and the coupling between them.  
The electron transmission coefficients $T_{\alpha,\beta}$ between 
different leads (indexed by $\alpha,\beta$) can be  expressed in terms 
of the retarded Green function, which has to be calculated explicitly. 
Once $T_{\alpha,\beta}$ are known, the transverse (Hall) $R_H$  resistance 
is given immediately by the formalism. The technical details of the 
calculation can be found in extenso in \cite{ONA}, for instance, and will 
not be repeated here.

The Hall transport is studied in the Chern insulating phase 
and  the  semimetallic phase, which have  been  noticed in Fig.9a.
The non-vanishing Hall conductance  results  from the asymmetry
of the transmission coefficients 
$T_{\alpha,\alpha+1} \ne  T_{\alpha+1,\alpha}$,
induced in its turn by the  phase  $\gamma$ that  breaks the TR symmetry.

In the energy range occupied by the  edge states, the numerical calculation 
indicates that the  asymmetry is total, i.e.,$T_{\alpha,\alpha+1}=1, 
T_{\alpha+1,\alpha}=0$, meaning that the  current is carried 
in a given  direction  (imposed by the sign of $\gamma$). 
This fact expresses the chirality of the edge states and the lack 
of backscattering in the transport process for the model under consideration. 
Combined with the observation that all the  other transmission coefficients 
$T_{\alpha,\beta}$ vanish, in line with the Landauer-B\"{u}ttiker theory, 
the consequence is  a {\it quantized} Hall conductance. 
In Haldane's terminology, this quantization is {\it anomalous}, as it occurs 
in the absence of  the external magnetic field. 

The numerical results supporting the above statements are shown in Fig.11, 
obtained under the condition $\Delta/4t_2 < 1$, which, 
as already shown in Fig.9 , induces  edge states in the insulating phase. 
The calculation is performed for a plaquette with all the four edges of 
the AB-type.  
The arrangement of the leads in the Hall device is sketched in Fig.11(right).
A gate potential  can be applied on the plaquette allowing the scanning 
all energies $E$ in the spectrum. 

As evident in Fig.11, as long as $\gamma$ crosses the gap, the transmission 
coefficient $T_{21}$ vanishes in the left gap, but manifests perfect 
transmission ($T_{21}=1$) in the right one. Since the transmission $T_{12}$ 
behaves vice-versa, one concludes that the edge states exhibit opposite
chirality in the two sectors of the 'binocular' spectrum. 
Accordingly, the Hall resistance is $R_H=\pm h/e^2$, the sign being different 
in the two sectors. 
In Fig.11, the blue curve corresponds to $E=0$, the red curve  to $E=1$, 
the difference in appearance  coming from the different ranges occupied
(on the $\gamma$-axis) by the edge states  at the two energies 
(one may check with Fig.9b).

The next task is to prove the robustness  of edge states against
disorder, which is the fingerprint of their topological character. 
The topic is addressed in what follows in conjunction with the
transport properties of  the semimetal.
In the semimetallic phase,  reached  under the conditions 
$\Delta/4t_2 = 1$ and $\gamma\ne0$, the  calculation of the 
Hall resistance  reveals a new conceptual  aspect. It refers to the 
disorder-driven AQHE arising in the semimetallic phase. 

Similar to the situation  met in Fig.8 for the ribbon geometry,  
the energy spectrum of the confined system in the semimetallic phase 
exhibits  edge states intercalated among the bulk states.
However, any perturbation, including the plaquette-lead coupling,
affects the spectral properties, giving rise to the  hybridization 
of the two types of states.
This yields backscattering and impedes the quantization of the
transport response. With the purpose of reducing the hybridization,
we introduce disorder, which induces localization and hence reduces 
the superposition of the two kinds of states.

%Fig11
\begin{figure}
\hskip-1.5cm
	\includegraphics[angle=-0,width=0.28\textwidth]{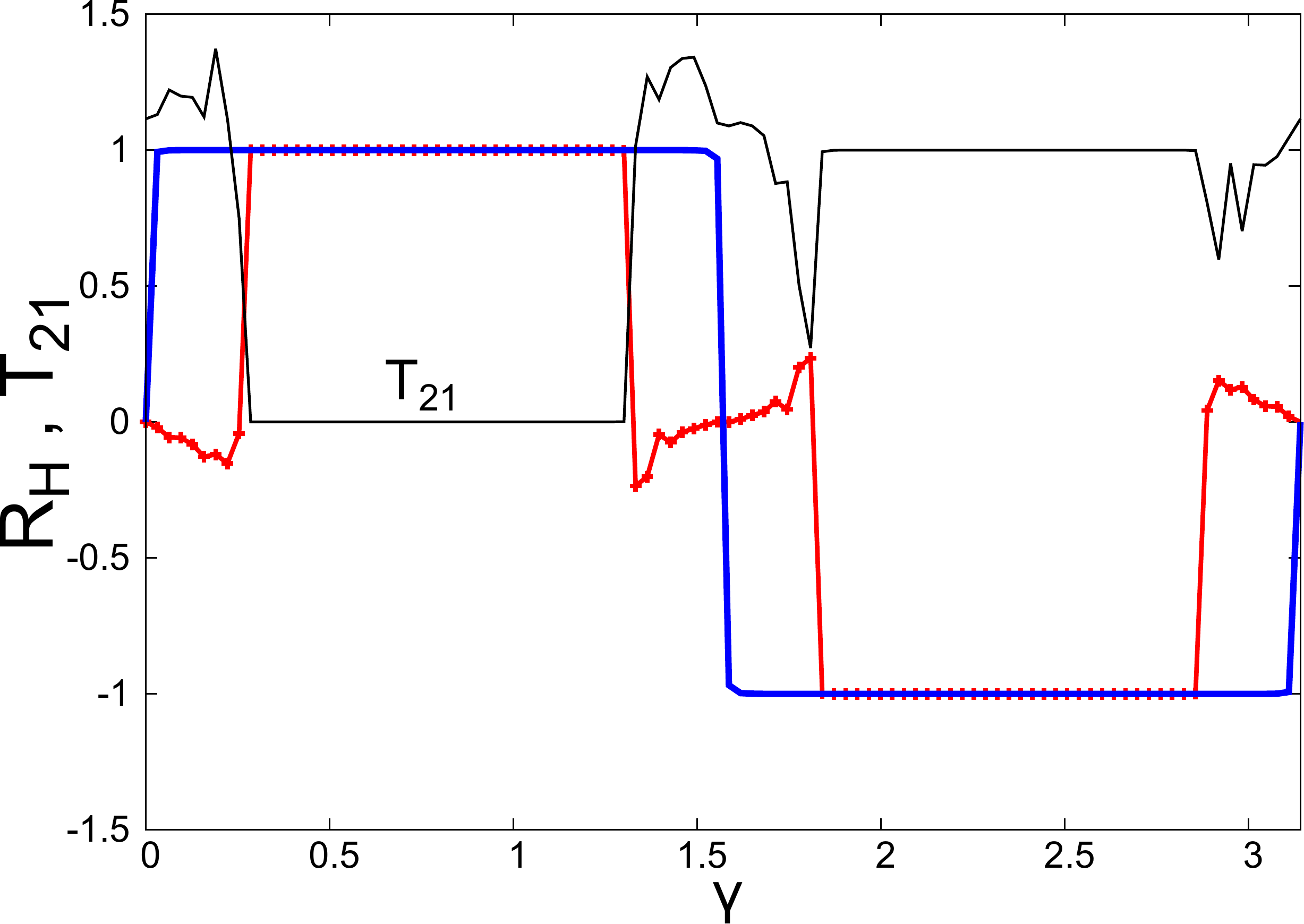}
\vskip-2.7cm
\hskip6.0cm
	\includegraphics[angle=-0,width=0.12\textwidth]{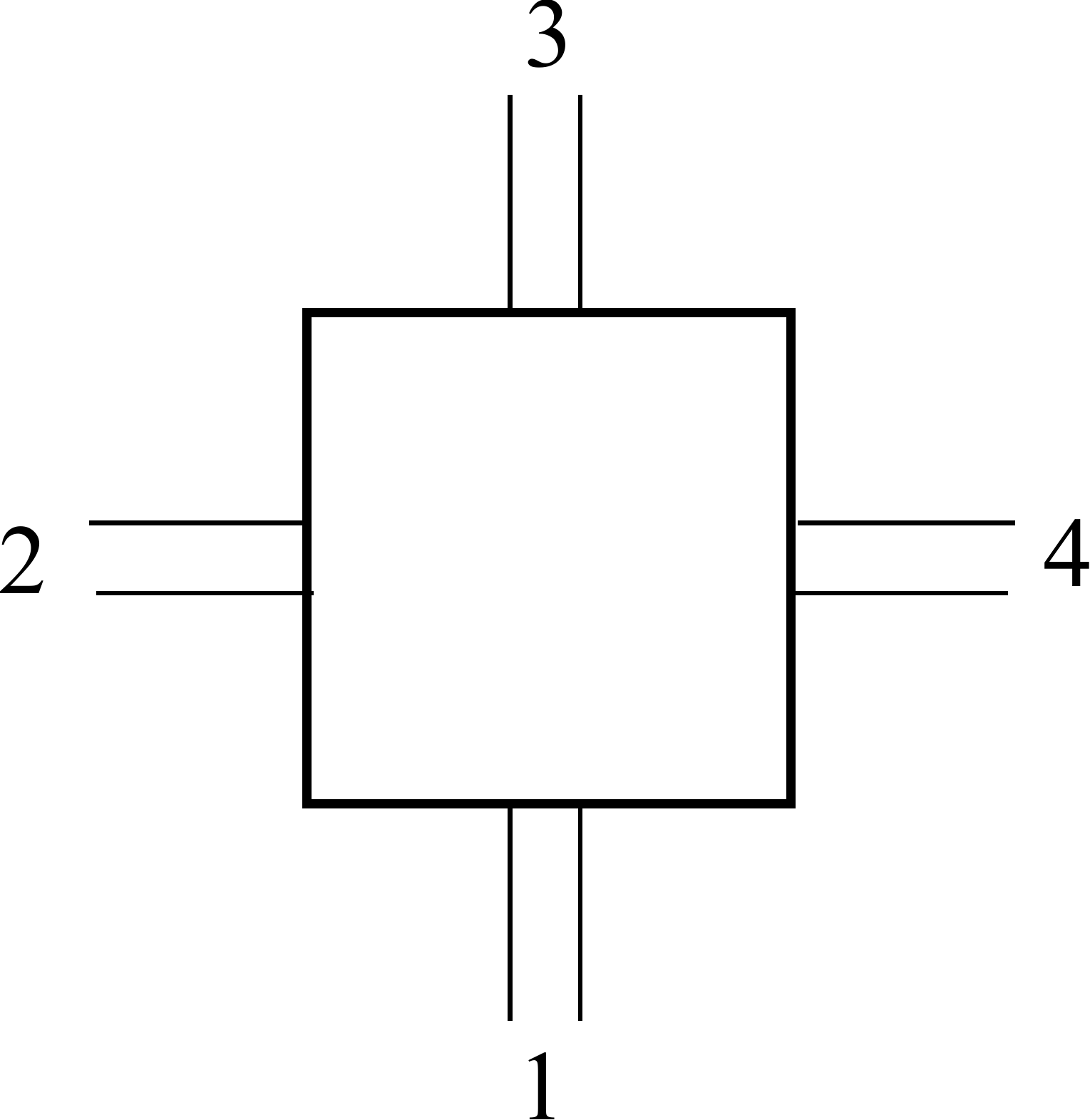}
\vskip0.7cm
\caption{(Color online) Hall resistance  versus the 'local flux' 
$\gamma$ in the Chern insulating phase ($t_2=1,\Delta=0$)
for two values of the Fermi level $E_F=0$ (blue) and $E_F=1$ (red).
As an example, transmission $T_{21}$ is also
shown for $E_F=1$. The quantization is manifest in the range occupied by 
the edge states (check with Fig.9b). (Calculation is performed for a 
25$\times$25 unit cells plaquette of AB type, with four leads 
attached in  crossed geometry as sketched on the right.) }
\end{figure}
%Fig12
\begin{figure}
        \includegraphics[angle=-0,width=0.23\textwidth]{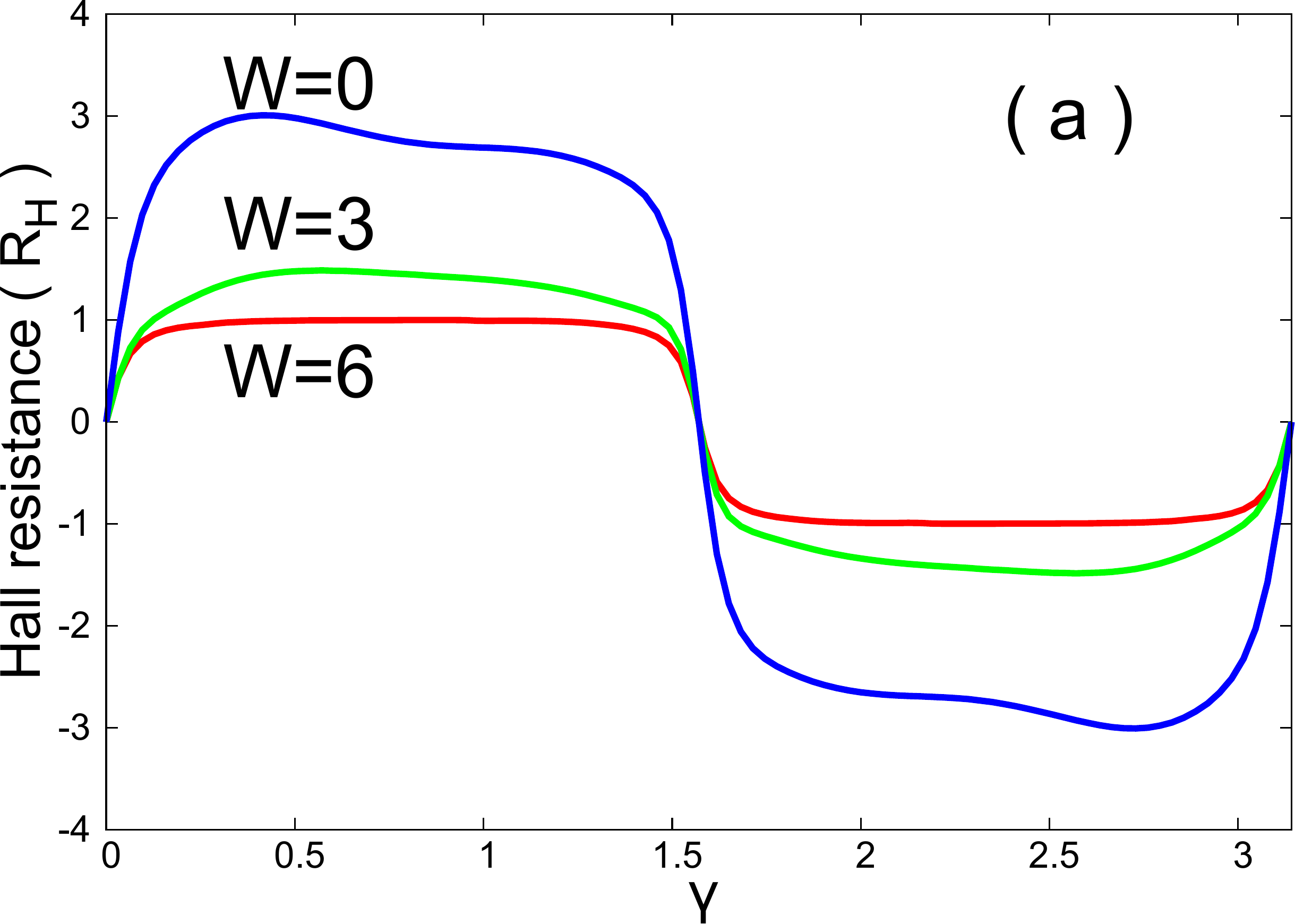}
        \includegraphics[angle=-0,width=0.23\textwidth]{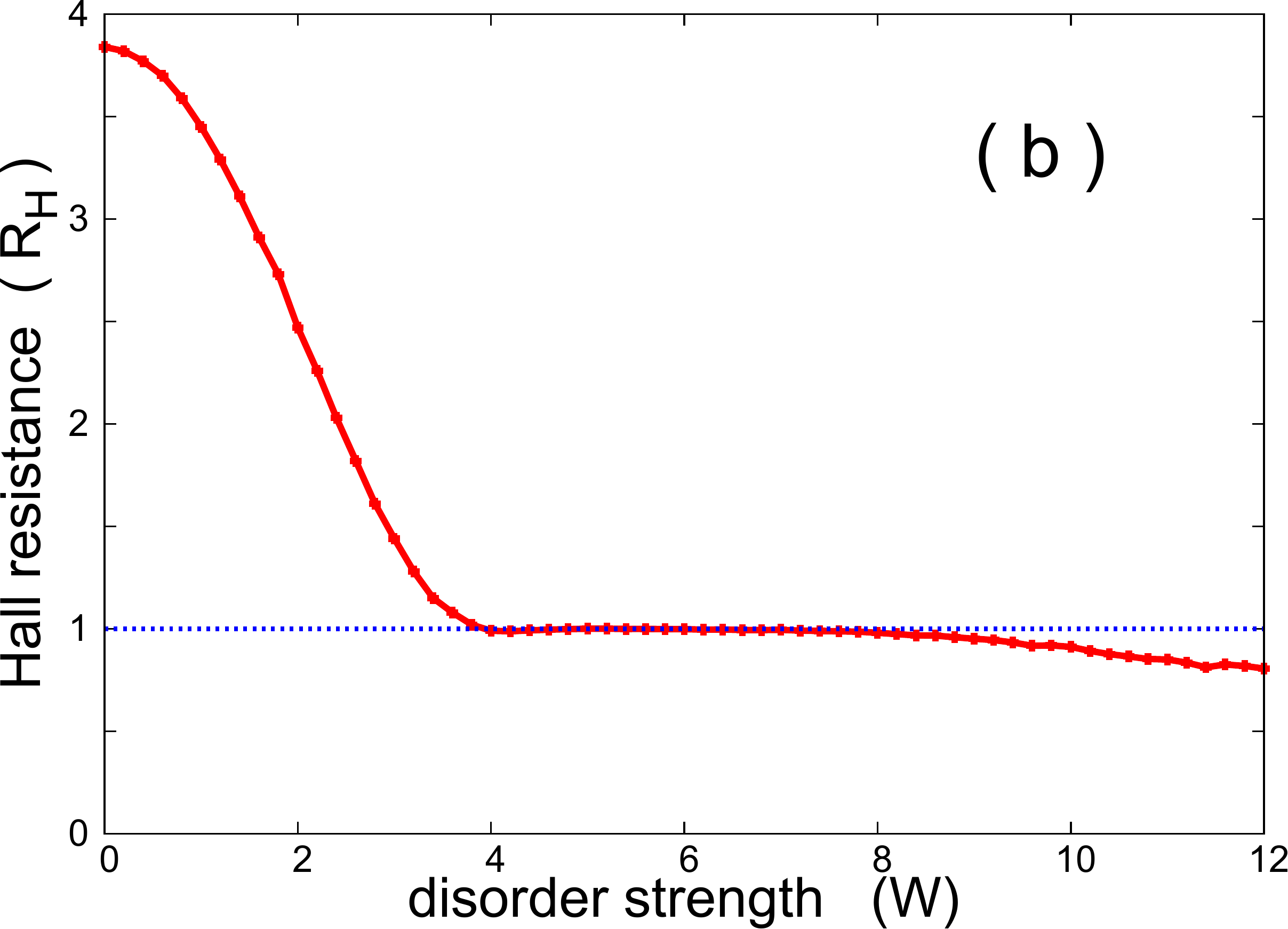}
\caption{(Color online) Disorder effect  on the Hall resistance 
in the semimetallic phase ($t_2=1, \Delta=4; E=0$). (a)  $R_H$ as function of  
$\gamma$ for different disorder strengths $W$. (b)  $R_H$  as function of 
$W$ at $\gamma=\pi/4$.   
The trend towards  resistance quantization with increasing disorder 
is observed  in both panels (plaquette dimension is 15$\times$15 unit cells, 
average is performed over 1000 disorder configurations; data are obtained for  
disordered AB-plaquette in panel (a), and   AA-plaquette in (b)).}
\end{figure}
According to the Anderson's localization theory \cite{Abrahams,Wolfle}, 
all quantum states in the {\it infinite} two-dimensional systems are 
localized, irrespective how weak the disorder is. 
On the other hand, in the finite-sample, the 1D states 
(the edge states in our case) should localize first, and  the 2D bulk 
states afterwards (The difference comes from the
different localization length $\lambda$ of the two types of states, 
which has to be compared to the typical dimension $L$ of the sample).
This should be true only if the 1D states under discussion 
are not protected against disorder. 
However, in our case, the  attention signal comes from the 
calculation  of the Hall resistance that reveals the tendency toward
quantization  $R_H=\pm 1 h/e^2$ as the disorder in the system is 
increased gradually.

In Fig.12a we show the  Hall resistance for different strengths
of the Anderson disorder characterized by the range [-W,W], in which the
diagonal energies of the tight-binding Hamiltonian are randomly distributed.
It is to  observe the evolution of the curve $R_H(\gamma)$ from  
the ordered case $W=0$ to the disordered one  $W=6$, which shows the 
quantum plateau (red curve). The dependence on  disorder can be  
remarked also in Fig.12b, where  $R_H$ is calculated as function of the 
disorder strength $W$ at fixed $\gamma=\pi/4$. 
The topological behavior $R_H=1$ is reached at $W\approx4$ 
(for the given plaquette dimensions), while at very strong disorder 
the Hall plateau is spoiled, as it should.

This outcome demonstrates that, in the disordered semimetal, 
the states in the bulk become localized  and 
do not contribute to the transport process, while the edge states 
are robust against disorder and give rise to the anomalous quantum Hall effect.

\section{conclusions}
In conclusion, in the frame of a confined diatomic lattice model with broken
time-reversal symmetry and hopping to the next-nearest-neighbors,
we have proved the occurrence of chiral edge states, robust to disorder, 
which may exist in the absence of an external magnetic field. 
The states are the physical support for the  anomalous quantum Hall effect.

The case of the  infinite lattice  allows the exact diagonalization 
of the Hamiltonian in the momentum space, such that different phases 
(semimetallic, conventional insulator or Chern insulator) can be identified 
from the expression of the energy spectrum and the Hamiltonian symmetries.
The spectral   properties  depend on the  set of parameters
$\gamma$ (which controls the time-reversal), $t_2$ (n.n.n.hopping),
and $\Delta$ (atomic energies staggering).
The electron-hole, time-reversal,  and inversion symmetries are discussed.
The diatomic lattice with n.n.n. hopping is an example of non-bipartite
lattice  that exhibits a symmetric spectrum around the zero energy.
The energy spectrum of the semimetallic phase may exhibit
one or two touching points, depending on the circumstences
discussed in Sec.II. 

The question under discussion is how the boundary conditions affect 
the spectral properties, and more precisely, whether the confinement 
may give rise to chiral edge states that  should be the support of the 
quantum anomalous Hall effect.  

In first instance,   the boundary effect is noticed in the 
ribbon (strip) geometry. We tailor the ribbons in two different ways,
obvious in Fig.1, which are called AA/BB and AB/AB ribbons, respectively.
The notable results are : i) the presence of the   
degenerated flat band composed by edge states in the AA/BB ribbon (Fig.4a),
ii) the occurrence in the semimetallic phase of edge states embedded in 
the continuum of the bulk states (Fig.5a and b), iii) the emergence of
chiral edge states in the insulating phase of  both ribbons if
$\Delta/4t_2<1$ (Fig.7a and b). 
All these situations are  presented in the phase diagram Fig.8.

The finite size plaquette geometry allows for the calculation of the 
transport properties and disorder effects. The plaquette has a rectangular 
shape, and can be obtained by cutting the ribbon perpendicularly. 
One has to pay attention to the requirement of zero total flux through 
the plaquette. By attaching four semi-infinite leads one simulates 
a Hall device, and the Hall resistance is calculated numerically 
following the Landauer-B\"{u}ttiker recipe.

The energy spectrum of the isolated plaquette as function of the
phase $\gamma$ shows a 'binocular' structure, whose gap is filled with
edge states if $\Delta/4t_2<1$ (Fig.9), and is empty in the opposite case.
By calculating the Hall resistance, it turns out that the edge states 
generated by the model are chiral and responsible for the anomalous 
quantum Hall effect (Fig.11). 

The semimetalic phase, occurring under conditions $\gamma\ne0$ and
$\Delta/4t_2=1$, deserves special attention. 
Being characterized by intercalated edge and bulk states, 
it cannot exhibit quantized transport as long as the bulk states are  
carrying  electrons.  By using Anderson disorder, we succeed to localize 
the bulk states and, at the same time, to observe a quantized  
Hall resistance at sufficiently strong  disorder (Fig.12).
This says that the edge states we have evidentiated in the semimetallic phase 
are chiral and robust against disorder. 
We call this effect as {\it disorder-driven} anomalous quantum Hall effect.

\section{Acknowledgments}
We acknowledge financial support from the Core Program PN18-11 and 
grant PNIII-P4-ID-PCE-2016-0084 of the  Romanian Ministry of Research 
and Innovation. We are much indebted to M. Tolea for useful discussions.


\begin{thebibliography}{30}
\bibitem{Note}
We shall see in Fig.9b that inside the  gap  $dE/d\gamma$
changes its sign, but Fig.11 says that the Hall effect keeps a given sign. 
This says that
the  derivative of the energy with respect to the phase $\gamma$ 
cannot be identified with the chirality of the edge state.
\end{thebibliography}
\end{document}